\documentclass{aa}  

\usepackage{graphicx}
\usepackage{txfonts}
\usepackage{float}
\usepackage{enumitem}
\usepackage{ctable}
\begin{document} 

   \title{Mass determination of protoplanetary disks from dust evolution}

   \author{Riccardo Franceschi \inst{1}, Tilman Birnstiel\inst{2,3}, Thomas Henning\inst{1}, Paola Pinilla \inst{1,4}, Dmitry Semenov\inst{1,5}, Apostolos Zormpas\inst{2}
          }

   \institute{Max-Planck-Institut f\"ur Astronomie (MPIA),
              K\"onigstuhl 17,  69117 Heidelberg, Germany
              \and University Observatory, Faculty of Physics, Ludwig Maximilians University, Scheinerstr. 1, 81679 Munich, Germany
              \and Exzellenzcluster ORIGINS, Boltzmannstr. 2, D-85748 Garching, Germany 
              \and Mullard Space Science Laboratory, University College London, Holmbury St Mary, Dorking, Surrey RH5 6NT, UK
              \and Department of Chemistry, Ludwig Maximilian University, Butenandtstr. 5-13, D-81377 Munich, Germany\\
              \email{franceschi@mpia.de}
              }

 
  \abstract
   {The mass of protoplanetary disks is arguably one of their most important quantities shaping their evolution toward planetary systems, but it remains a challenge to determine this quantity. Using the high spatial resolution now available on telescopes such as the Atacama Large Millimeter/submillimeter Array (ALMA), recent studies derived a relation between the disk surface density and the location of the "dust lines". This is a new concept in the field, linking the disk size at different continuum wavelengths with the radial distribution of grain populations of different sizes.}
   {We aim to use a dust evolution model to test the dependence of the dust line location on disk gas mass. In particular, we are interested in the reliability of the method for disks showing radial substructures, as recent high-resolution observations revealed.}
   {We performed dust evolution calculations, which included perturbations to the gas surface density with different amplitudes at different radii, to investigate their effect on the global drift timescale of dust grains. These models were then used to calibrate the relation between the dust grain drift timescale and the disk gas mass. We investigated under which condition the dust line location is a good mass estimator and tested how different stellar and disk properties (disk mass, stellar mass, disk age, and dust-to-gas ratio) affect the dust line properties. Finally, we show the applicability of this method to disks such as TW Hya and AS 209 that have been observed at high angular resolution with ALMA and show pronounced disk structures.}
   {Our models without pressure bumps confirm a strong dependence of the dust line location on the disk gas mass and its applicability as a reliable mass estimator. The other disk properties do not significantly affect the dust line location, except for the age of the system, which is the major source of uncertainty for this mass estimator. A population of synthetic disks was used to calibrate an analytic relation between the dust line location and the disk mass for smooth disks, finding that previous mass estimates based on dust lines overestimate disk masses by about one order of magnitude. Radial pressure bumps can alter the location of the dust line by up to $\sim 10$ au, while its location is mainly determined by the disk mass. Therefore, an accurate mass estimation requires a proper evaluation of the effect of bumps. However, when radial substructures act as traps for dust grains, the relation between the dust line location and disk mass becomes weaker, and other mass estimators need to be adopted.}
   {Our models show that the determination of the dust line location is a promising approach to the mass estimate of protoplanetay disks, but the exact relation between the dust line location and disk mass depends on the structure of the particular disk. We calibrated the relation for disks without evidence of radial structures, while for more complex structures we ran a simple dust evolution model. However, this method fails when there is evidence of strong dust traps. It is possible to reveal when dust evolution is dominated by traps, providing the necessary information for when the method should be applied with caution.}

    
   \authorrunning{R. Franceschi \inst{1}, T. Birnstiel \inst{2,3}, Th. Henning \inst{1}, P. Pinilla \inst{1}, D. Semenov \inst{1,4}
          }

   \institute{Max-Planck-Institut f\"ur Astronomie (MPIA),
              K\"onigstuhl 17,  69117 Heidelberg, Germany
              \and University Observatory, Faculty of Physics, Ludwig Maximilians University, Scheinerstr. 1, 81679 Munich, Germany
              \and Exzellenzcluster ORIGINS, Boltzmannstr. 2, D-85748 Garching, Germany 
              \and Mullard Space Science Laboratory, University College London, Holmbury St Mary, Dorking, Surrey RH5 6NT, UK
              \and Department of Chemistry, Ludwig Maximilian University, Butenandtstr. 5-13, D-81377 Munich, Germany\\
              \email{franceschi@mpia.de}
              }
   \titlerunning{PPDs mass determination from grain evolution}
   \authorrunning{Franceschi et al.}
   \maketitle
%

\section{Introduction}

The total mass of a protoplanetary disk is of prime importance for disk evolution and planet formation studies (e.g., \citealt{Mordasini12, Birnstiel16}). Specifically, the disk mass is fundamental for pebble accretion models. As an example, \citealt{Lambrechts14} show that the disk mass must ensure a high enough radial flux of pebbles (millimeter to centimeter sizes) so the core of that giant planet may form before the disk is depleted from gas. The disk mass cannot be directly determined as most of its mass resides in H$_2$, which is a symmetric molecule without a dipole moment and millimeter rotational transitions. The disk surface density distribution is traditionally inferred from dust continuum emission or molecular tracers, such as HD and CO (e.g., \citealt{Bergin13, Ansdell16, Long17, Bergin18, Kama20}).\par
The dust surface density is measured from its thermal emission at (sub)millimeter wavelengths, assuming it is optically thin, and converted to a total surface density through an assumed dust-to-gas ratio (e.g., \citealt{Andrews05, Andrews07}). The drawback in using this approach is that the dust-to-gas ratio is not well constrained in the disk environment, and it can be significantly different from the canonical interstellar value $10^{-2}$ \citep{Williams11} due to dust evolution. To estimate the surface density of solid material, opacity must also be assumed, which is an additional source of uncertainty (e.g., \citealt{Henning96, Andrews05, Birnstiel18}). Moreover, continuum observations do not trace emission from grains that are much larger or much smaller than the observed wavelength. It is therefore possible that dust mass estimates are missing a significant fraction of the total dust mass. This is even more true if part of the emission is optically thick \citep{Liu19, Zhu19}.\par

The emission from CO rotational transitions of one or more of its optically thin isotopologues can also be used to estimate the gaseous content of a disk via an assumed CO-to-H$_2$ ratio (e.g., \citealt{Schwarz16, Zhang17, Booth19}). However, this conversion factor is quite uncertain, as there is mounting evidence that CO is depleted from the gas phase through freeze out, isotope-selective photodissociation and conversion to more complex chemical species (e.g., \citealt{vanZadelhoff01,Miotello16, Schwarz16}), and the typical assumed CO abundance of 10$^{-4}$ based on interstellar estimates has been questioned.\par

HD, on the other hand, is a more direct tracer of molecular hydrogen, and the derived disk mass from HD measurements for TW Hya is higher by ~2 orders of magnitude than CO estimates, due to CO depletion from the gas phase \citep{Bergin13, McClure16, Schwarz16, Kama20}. However, HD emission only comes from warm regions in the disk ($T \sim 30-50 \; K$) \citep{Bergin13, Trapman17} and can therefore only provide lower limits to the gas mass. Moreover, HD observations only exist for TW Hya \citep{Bergin13}, GM Aur and DM Tau \citep{McClure16}, and there have not been new HD observations after the \textit{Herschel} mission has ended.\par

Given the difficulties in converting observable disk quantities to a total gas mass, in this paper we used theoretical models to test the novel approach to disk mass measurements introduced in \cite{Powell17, Powell19}. This method links the maximum radial extent of the emission at a given wavelength, the so-called dust line, to the local surface density of the disk. Previous theoretical studies \citep{Birnstiel12, Birnstiel14} demonstrate that dust evolution is driven by grain growth and fragmentation. These processes are regulated by drift in the outer disk and cause segregation of the radial distribution of grains with their size. In \cite{Powell17, Powell19}, multiwavelength observations of dust continuum emission have been used to derive the grain size distribution and to estimate the gas surface density distribution for a sample of protoplanetary disks.\par

In this paper, we tested the theoretical foundation of this approach through numerical modeling in order to take all grain evolution processes into account. The model is described in Section \ref{Sec:model}, and in Section \ref{Sec:dustline} we discuss the definition of dust lines and how these can be used to estimate the disk surface density. We also investigate under which conditions dust trapping will limit the applicability of dust lines for the estimates of disk masses. In Section \ref{Sec:params} we study the model dependence on the physical parameters. Once we verify the goodness of the dust line as a mass estimator, we calibrate it using a population of synthetic disks in Section \ref{sec:population}. Next, we proceed to study how deviation from a smooth gas distribution may affect a mass estimate in Section \ref{sec:structures}. Finally, in Section \ref{Sec:disks} we apply it to individual disks and compare our results with the observationally derived results in \cite{Powell17, Powell19}.

\section{Methodology}
\label{Sec:model}

\cite{Powell17, Powell19} show how considerations about drift and growth timescales can be used to estimate the gas surface density at the outer edge of the emitting region of any given dust grain population (they call this location the \textit{dust line}). Our aim is to test and supplement this idea with dust evolution models. To this purpose we briefly summarize their method.\par

In the outer regions of evolved disks, particle drift dominates both grain growth and collisional fragmentation. It determines the grain size distribution and, in particular, the maximum grain size at any given radial location. In this simplified scenario, grain evolution happens in two regimes: a fragmentation regime in the inner disk and a drift regime in the outer disk. Grains in the outer disk have too low relative velocities for fragmentation to affect their evolution, and their maximum size at any given radius is determined by the drift. In the inner disk, however, the relative velocities are higher, and grain evolution is fragmentation limited \citep{Birnstiel12, Birnstiel14}.\par

The timescale on which particles grow (e-folding timescale) can be estimated to be \citep{Brauer08, Birnstiel12}:

\begin{equation}
    \label{eq:growth_timescale}
    t_{growth} = \frac{1}{\epsilon \; \Omega_K},
\end{equation}

where $\epsilon$ is the local dust-to-gas ratio and $\Omega_K$ the Keplerian angular velocity, although the initial stage of this growth might be longer \citep{Powell19}. If particle growth is not halted by collisional effects such as fragmentation or bouncing \citep{Guttler10, Zsom10, Birnstiel12}, then the radial migration limits the grain particle sizes, since large grains migrate faster than growth can replenish them. The largest achievable grain size, called the drift limit, is therefore given by equating the drift and growth timescales \citep{Birnstiel12}, where the drift timescale is:

\begin{equation}
    \label{eq:drift_timescale}
    t_{drift} = \frac{r}{v_{drift}},
\end{equation}

and $v_{drift}$ is the radial drift velocity \citep{Whipple72}. The radial drift occurs because of the head-wind from the subkeplerian gas flows against the Keplerian grains, that lose angular momentum. The drift velocity is \citep{Weidenschilling77}:

\begin{equation}
    v_{drift} = - \frac{2 \Delta v}{\mathit{St} + \mathit{St}^{-1}},
\end{equation}

where $\Delta v$ is the difference between the gas and the grains orbital velocity, and $\mathit{St}$ is the Stokes number, the dimensionless ratio of stopping time to the dynamical timescale $\mathit{St} = t_{stop} \; \Omega_K$ (which depends on the grain size). If particle growth and drift have had enough time to proceed, then particles that drift on timescales shorter than the age of the disk should not exist. In other words, particles of a given size need to be located in regions where their drift timescale is comparable or longer than the disk age. The sharp edges observed in millimeter emission of a number of disks  \citep[e.g.,][]{Perez15, Andrews16, Tazzari16} are therefore interpreted by \cite{Powell17} as separating the well coupled ($t_{drift} > t_{disk}$) and significantly decoupled grains ($t_{drift} < t_{disk}$). Since in the drift limit ($t = t_{drift}$) the maximum grain size at a given location is directly proportional to the local dust surface density $\Sigma_{d}$ \citep{Birnstiel12}, these sharp edges can be used to link the maximum grain size to $\Sigma_d$. Larger particles have smaller drift timescales, meaning that the dust lines will evolve with disk age, and we expect the location of the dust lines to decrease with the disk age. This motivates \cite{Powell17} to equate the above timescales to the disk age at the position of the dust outer edge. To the additional constraint of equating the timescales at the dust line location to also the disk age gives us a relation for the dust-to-gas ratio appearing in Eq.\ref{eq:growth_timescale}:

\begin{equation}
    \label{eq:d2g_ratio}
    \epsilon = \frac{1}{t_{disk} \; \Omega_K}.
\end{equation}

This allowed us to constrain both the dust and gas surface densities if the particle size is known. If continuum observations are assumed to be dominated by grains of size $a = \lambda / 2\pi$, according to the Mie scattering theory, then multiwavelength observations that show dust lines at different positions can be used to reconstruct the surface densities of dust and gas and thus one can derive mass estimates of the solids and the gas that do not depend on opacities, but are based on dynamical considerations (i.e., drift speeds and collisional times).\par

Smaller particles have a longer drift timescale, and their emission will come from a broader region, while for larger grains the emission region will shrink. Following the derivation in \cite{Powell17}, this condition leads to an expression for the gas surface density at the edge of a given dust population emitting region, the so-called \textit{dust line location}:

\begin{equation}
    \label{eq:sigma_g}
    \Sigma(r) = \frac{t_{disk} v_0 \rho_s a}{r},
\end{equation}

where $v_0$ is the drift velocity of the fastest-drifting grains (with $\mathit{St} = 1$), $\rho_s$ the internal density of grains (taken to be 1.6 g/cm$^3$) and $r$ is the maximum radius where grains of size $a$ can be found. If multiwavelength observations of the continuum emission are available, the size of each one of these emitting regions, assuming an emitting grain population of size $\lambda_{obs} / 2 \pi$, can be put in Eq.\ref{eq:sigma_g} to estimate the gas surface density at the dust line location. These estimates can then be used to fit theoretical models to the gas surface density distribution, such as the Lynden-Bell-Pringle self-similar solution. This approach has been used to estimate the disk mass of TW Hya \citep{Powell17} and the masses of a sample of other disks \citep{Powell19} from dust continuum observations.\par

To test this method, we modeled grain evolution using the two population model described in \cite{Birnstiel12}. The dust is evolving in a viscous environment by considering two populations of grains: a small grain population which remains well coupled to the gas structure, with a size of 0.1 $\mu$m, and a large grain population which may grow and drift (that carry most of the mass). This model is ideal for our purpose as it is computationally inexpensive and it is calibrated to match the grain size and mass flux in more comprehensive full population models.\par

Modeled disks were initially populated with micron-sized particles with a dust-to-gas ratio 1/100. Particles can stick and grow when colliding if their relative velocity is below a threshold, set to 10 m/s in our model, based on numerical and laboratory experiments (e.g., \citealt{Gundlach11, Wada09}). Dust dynamics is dominated by drift and turbulent diffusion. Dust diffusion is assumed to be the same as the turbulent gas viscosity \citep{Youdin07} and the turbulent velocities are proportional to $\sqrt{\alpha}$ \citep{Ormel07}, where $\alpha$ is the disk effective viscosity parameter \citep{Shakura73}.\par

We modeled the disk evolution through gas viscosity, growth and radial drift of solid material. The gas structure was assumed to follow the \cite{Lynden-Bell74} self-similar solution, following from mass and momentum conservation:
\begin{equation}
\label{eq:LBP}
    \Sigma(r) = \Sigma_c \left(  \frac{r}{r_c} \right)^{-\gamma} \exp \left[ -\left(  \frac{r}{r_c} \right)^{2-\gamma} \right].
\end{equation}

Following \citet{Birnstiel12}, at each radius, we evolved the small and large grain populations. The maximum grain size was set to the lowest value between the drift limit or the fragmentation limit, or was limited by the growth time if particles had not yet grown to one of the size limits. Fragmentation of dust particles stops further growth because the relative velocity of grains increases with their Stokes number. When they reach the fragmentation threshold velocity, particle collisions destroys the grains instead of sticking them together to form larger grains. Since the grains' relative velocities due to turbulence increase with grain size, the maximum grain size in the fragmentation limited regime can be estimated as:

\begin{equation}
    a_{frag} = f_r \frac{2}{3 \pi} \frac{\Sigma_g}{\rho_s \alpha} \frac{u_f^2}{c_s^2},
\end{equation}

where $f_r \approx 0.37$ is a calibration factor obtained from full dust evolution codes \citep{Birnstiel12},  $\Sigma_g$ is the local gas surface density, $\rho_s$ is the grain internal density, $c_s$ is the gas sound speed and $u_f$ is the fragmentation velocity .\par

In the drift limited regime, large grains are removed when their drift timescale is shorter than the time required to form these  grains. In this regime, the grains drift at least as fast as they grow, and the maximum grain size is reached when the drift and growth timescales are equal:

\begin{equation}
    a_{drift} = f_d \frac{2 \Sigma_d}{\pi \rho_s} \frac{V_k^2}{c_s^2} \gamma^{-1},    
\end{equation}

where $f_d \approx 0.55$ is another calibration factor, $\Sigma_d$ is the dust surface density, $V_k$ is the Keplerian velocity, and $\gamma = d\log{P} / d\log{r}$.\par

The grains' relative velocities are determined by drift and turbulence. The temperature structure establishes the maximum turbulent velocity, determining where the disk is going to be drift- or fragmentation-dominated. The temperature of the dust is assumed to be a power-law that depends on the stellar luminosity $L_\star$:

\begin{equation}
    T_d(r) = T_{10} \left( \frac{r}{10 \; \mathrm{au}} \right)^{1/2} \left( \frac{L_\star}{L_\sun} \right)^{1/4},
\end{equation}

where $T_{10}$ is the temperature at 10 au, taken to be 30 K \citep{Andrews13, Tripathi17}.

\section{Dust line location}
\label{Sec:dustline}

\subsection{Dust line definition}
The dust line is defined as the outer edge of the disk observed at a specific (sub)millimeter wavelength. In some cases, where the continuum emission abruptly ends, this definition is straightforward. However in other cases, where the emission continuously decreases, this definition needs to be refined. From an observational point of view, the most straightforward definition would be an "emission" dust line, that is, the radius enclosing a fixed fraction of the disk's total luminosity at a specific wavelength. On the other hand, the physical quantity that concerns us is the dust mass distribution for a given grain size. The most convenient way would be then to define a "mass" dust line using the surface density distribution of the most emitting grains. However, it is not clear if these two possible definitions of dust line location overlap.\par

\begin{figure}
    \centering
    \includegraphics[width=8cm]{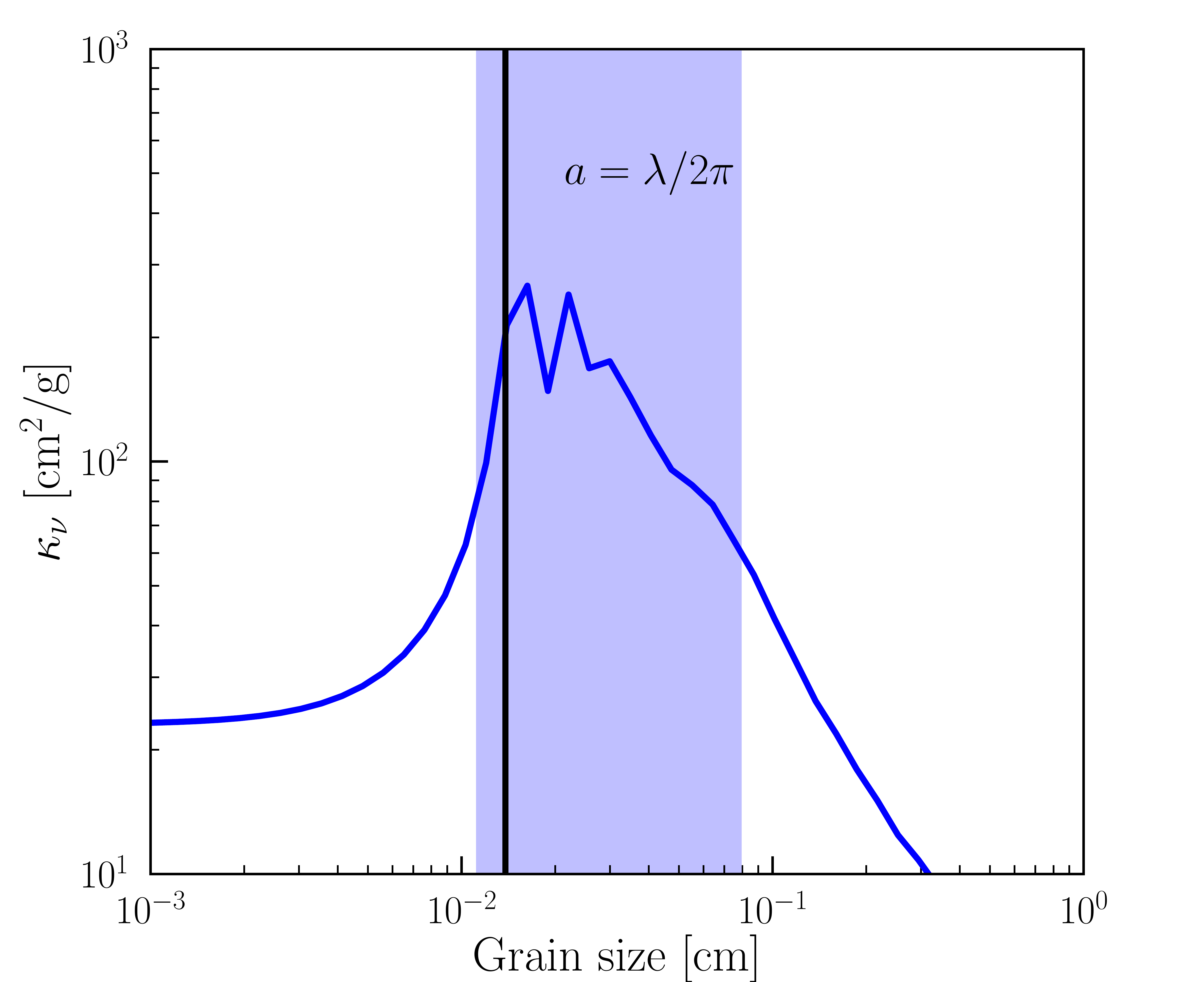}
    \caption{Dust opacity for a single grain population as a function of grain size at $\lambda =0.87$ mm \citep{Birnstiel18}. The shaded region represents the grain size range dominating the emission at 0.87 mm, the vertical black line indicates the size of grains dominating the emission according to the Mie scattering theory, with size $a = \lambda / 2 \pi$.}
    \label{Fig:opacity_0.87}
\end{figure}

The emission at a given wavelength $\lambda_{obs}$ is dominated by grains with characteristic size $\lambda_{obs} / 2 \pi$, and we can link the emission dust line to the mass dust line for the grain population $a = \lambda_{obs} / 2 \pi$. However, even if this grain population is dominating the emission, other grains are contributing as well. To study the link between dust populations and the emission dust line we take the example of the opacity curve for a single grain population as a function of the grain size at $\lambda_{obs} = 0.87$ mm in Fig.\ref{Fig:opacity_0.87} \citep{Ricci10}. The most emitting grains span an order of magnitude in grain size and have their opacity increased by at least an order of magnitude compared to other grain sizes. Their radial position can be reliably traced by a given percentage of the disk flux, but we still do not know how to relate this flux-defined dust line to a single dust population.\par

The shape of the opacity curve can be used to solve this degeneracy. The opacity curve presents a sharp drop in opacity around the characteristic size $\lambda_{obs} / 2 \pi$ where the maximum opacity is reached. Moving toward smaller grains, the opacity drops abruptly over a narrow range of grain sizes. This steep part of the opacity was termed the \textit{opacity cliff} by \citet{Rosotti20} and is highlighted in Fig.\ref{Fig:opacity_0.87}. An approximate solution to grain emission in the optically thin limit (and the Rayleigh–Jeans low-frequency approximation) is:

\begin{equation}
    \label{eq:RJ_limit}
    I_\nu = \frac{2 k_B T \nu^2}{c^2} \cdot (\kappa_\nu \; \Sigma_{d}),
\end{equation}

where $\nu$ is the observed frequency, $k_B$ is the Boltzmann constant, T is the dust temperature, $\kappa_\nu$ is the dust opacity at the observed wavelength, and $\Sigma_d$ is the dust surface density. In disks $T$, $\Sigma_d$ and maximum grain size drop with radius. Moving from the inner disk outward (where $2 \pi \, a \gg \lambda$) the opacity is initially increasing with radius, but then at the opacity cliff, it is quite abruptly dropping to a lower, roughly constant value. Outside of this drop, the dust suddenly has a lower opacity while $T$ and $\Sigma_d$ keep dropping, so there is little emission outside the opacity cliff. This initial increase and sudden drop of $\kappa(r)$ (due to the size sorting) is the feature that gives rise to the dust line. The grain size at the edge of the emission region is therefore the one at which the opacity cliff occurs, and this location corresponds to the edge of the surface density profile of this grain population. From an observational point of view, identifying the emission dust line at a wavelength $\lambda_{obs}$ with the mass dust line for the grain population $a = \lambda_{obs} / 2 \pi$ is a good approximation, and from now on we use the more generic term dust line. \par

Optical depth effects are another potential issue in the identification of the dust line location. If the dust emission is optically thin, it is straightforward to associate a steep drop in the brightness profile to the dust line. If the emission is instead optically thick, one could argue that the drop in density could be hidden in the optically thick region, and that the dust line is misplaced at the location where the emission goes from being optically thick to optically thin. Since the observed drop in the emission is sharp \citep{Birnstiel14, Andrews12, Powell17}, it is not expected to originate from opacity effects (that would show a much smoother transition). Indeed, observations show an intensity drop of about one order of magnitude over a very narrow radial range ($\Delta r/ r \lesssim 0.1$). As the dust line location falls within this radial range, we take $\Delta r/ r$ as the dust line relative error.\par

Given the previous considerations, we identify the dust line location as the radius enclosing $99 \%$ of the total dust emission, that is well representing the outer edge of the dust density profile in our simulations. The exact percentage of the total emission can vary, depending on the disk structure. This parameter can be tuned to match the location of the sharp intensity drop in the brightness profile, typical of the dust line. While other methods are possible, such as fitting the profiles with a broken power law or a similar function with a sharp cut-off, we found that the reliability of these methods can be impaired by the quality of the observations, and can be unreliable in disks showing radial substructures.\par

\subsection{Radial density bumps}
\label{sec:drift_regime}
As shown in Sec.\ref{Sec:model}, when the dust evolution is dominated by drift, the dust line location can be used as a proxy of the local disk gas surface density. However, observations show that disks often have features that can be interpreted as pressure bumps and dust traps (e.g., \citealt{Dullemond18, Huang18, Pinilla20}), suggesting that other physical processes could be determining the exact dust line location. To estimate the disk surface density distribution using dust lines we must first understand how drift is affected by these substructures. Indeed, bumps in the density profile create a positive pressure gradient slowing down the inward drift (e.g., \citealt{Pinilla12}). This can in turn move the dust lines to an outer radial position, and we must take this effect into account when estimating the disk surface density from the dust line location. If dust grains are efficiently trapped we would observe an increase in the emission at a range of (sub)millimeter wavelengths at the bump location, since millimeter-sized grains will be trapped. The width of this emission feature is going to change with wavelength, as large particles that are more affected by drift will be trapped more efficiently than small particles. With multiwavelength observations, we would be able to distinguish dust lines from traps. On the other hand, if the bump is not efficiently trapping the grains, it will only slow down the drift of the grains, and the dust line will not be affected.\par

\subsection{Fragmentation velocity}
In our simulations, we assumed a fragmentation threshold of 10~m/s, as discussed in Sec.\ref{Sec:model}. However, recent studies suggest lower thresholds to be more realistic. For instance, \cite{Musiolik19} demonstrates that even ices have no advantage over silicates in collisional growth processes, and the lower grain stickiness can lower the fragmentation threshold down to $\sim 1$~m/s.\par
To test how the fragmentation threshold affects our results, we ran our simulations with a lower threshold of 1~m/s. Our approach to the disk mass estimate requires the dust lines to lay in the outer disk, where dust evolution is drift dominated. By increasing the fragmentation threshold, we increase the size of the fragmentation limited inner disk, where our method cannot be applied. At low fragmentation velocities, it may be necessary to verify if the dust lines of larger grains are determined by drift or fragmentation. This can be done by checking that these dust lines do not gather at the same radial location. However, when the dust lines are drift dominated, the fragmentation threshold has no effect on the mass estimate. This is demonstrated by our models, which show no difference in the dust line location in the outer disk for disks of the same mass and a fragmentation threshold of 1~m/s and 10~m/s

\section{Simulation results}
\label{Sec:params}

In this section we show our model results and explore how different disk parameters affect the location of the dust line. The parameters taken into account are the disk mass $M_{disk}$, the stellar mass $M_{star}$, the age of the disk $t_{disk}$ and the initial dust-to-gas mass ratio $\epsilon$. As we are studying the evolution of large grains that are found in the disk mid-plane, we assumed the emission coming from vertical slabs of constant temperature and opacity (razor-thin disk). The brightness profile is then given by the simple radiative transfer equation:

\begin{equation}
    I_\nu = B_\nu (T) \left(  1  - e^{-\tau_\nu} \right),
\end{equation}

where $B_\nu(T)$ is the Planck function and $\tau_\nu$ the optical depth at a frequency $\nu$.\par

\begin{figure}
    \centering
    \includegraphics[width=8cm]{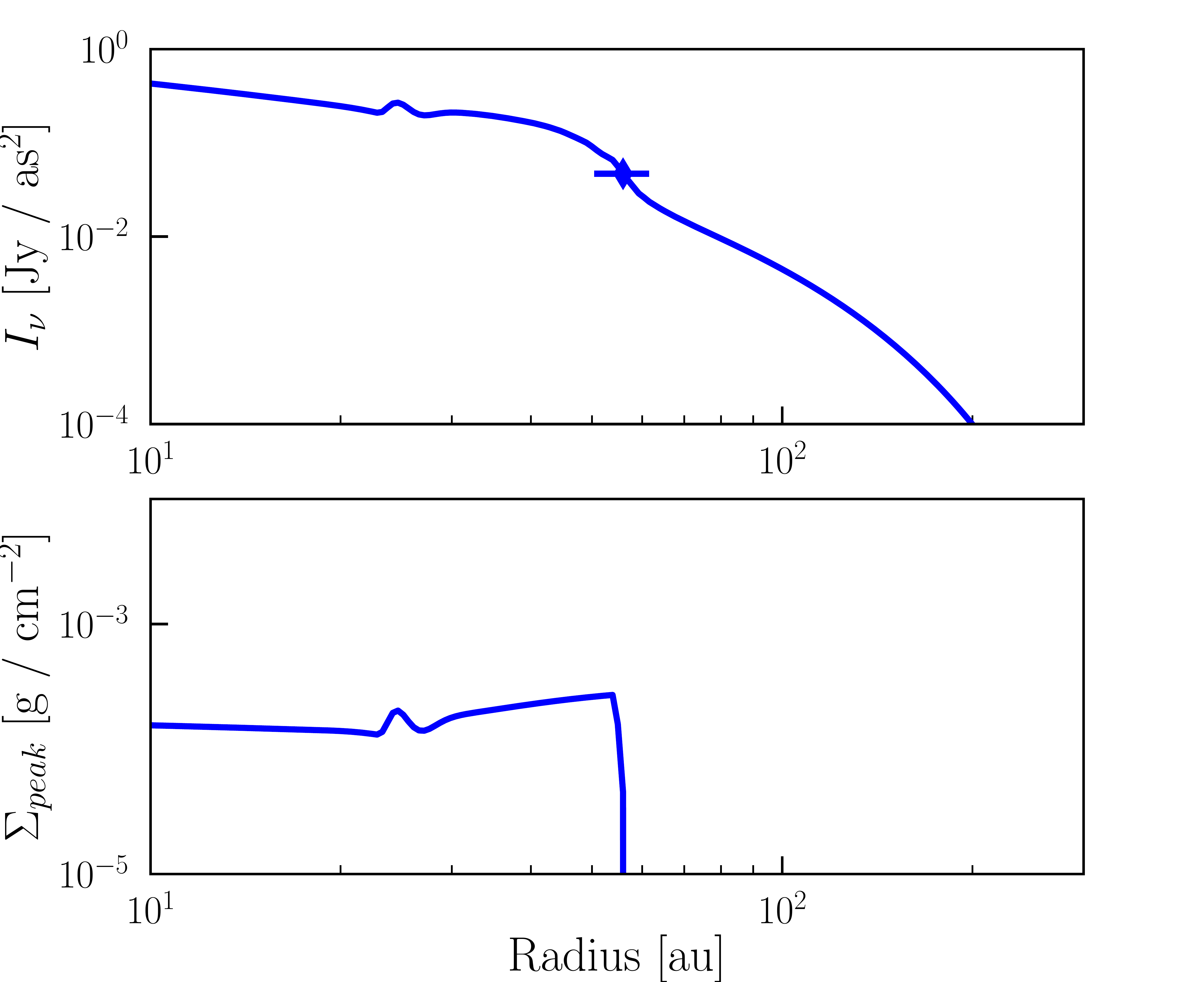}
    \caption{Reference model for the parameter study. The upper plot shows the dust emission profile at 0.87 mm, as predicted by the model. The dust line location, as defined in the text, is highlighted on the profile. The lower plot shows the surface density distribution of the most emitting grains. The dust line location in the first plot matches the outer edge of the density distribution in the lower plot for the same disk mass}
    \label{Fig:reference_model}
\end{figure}

The parameter study is based on a reference model of a disk with a column density $\Sigma_c = 175$ g/cm$^2$, corresponding to a disk mass of 0.11 $M_\sun$. For each parameter explored, we fixed the other parameters at the values used in the reference model, shown in Tab.\ref{table:reference_model}. The plots in Fig.\ref{Fig:reference_model} show how the dust emission profiles compare to the surface density distribution of the most emitting grains. It is clear from the plot that the dust line location represents the outer edge of the dust surface density distribution quite well.\par

\begin{table}
\label{table:reference_model}      
\centering                          
\begin{tabular}{c c}        
\hline\hline                 
Parameter & Value \\    
\hline                        
   $M_{disk}$ & 0.11 M$_\sun$ \\
   $r_c$ & $30$ au\\
   $\epsilon_0$ & $0.01$\\   
   $M_*$ & 0.8 M$_\sun$\\
   $L_*$ & 0.28 L$_\sun$\\
   $t_{disk}$ & 5 Myr\\
\hline                                   
\end{tabular}
\caption{Physical parameters for the reference model.}
\end{table}


Lastly, we discuss the effect of assuming a different gas surface density profile. As previously discussed, if the presence of substructure dominates the dust evolution, the dependence of the size of the emission region on the disk mass could be lost. The distinction between the two regimes is essential for the reliability of this technique.

\subsection*{Disk mass}
\begin{figure}
    \centering
    \includegraphics[width=8cm]{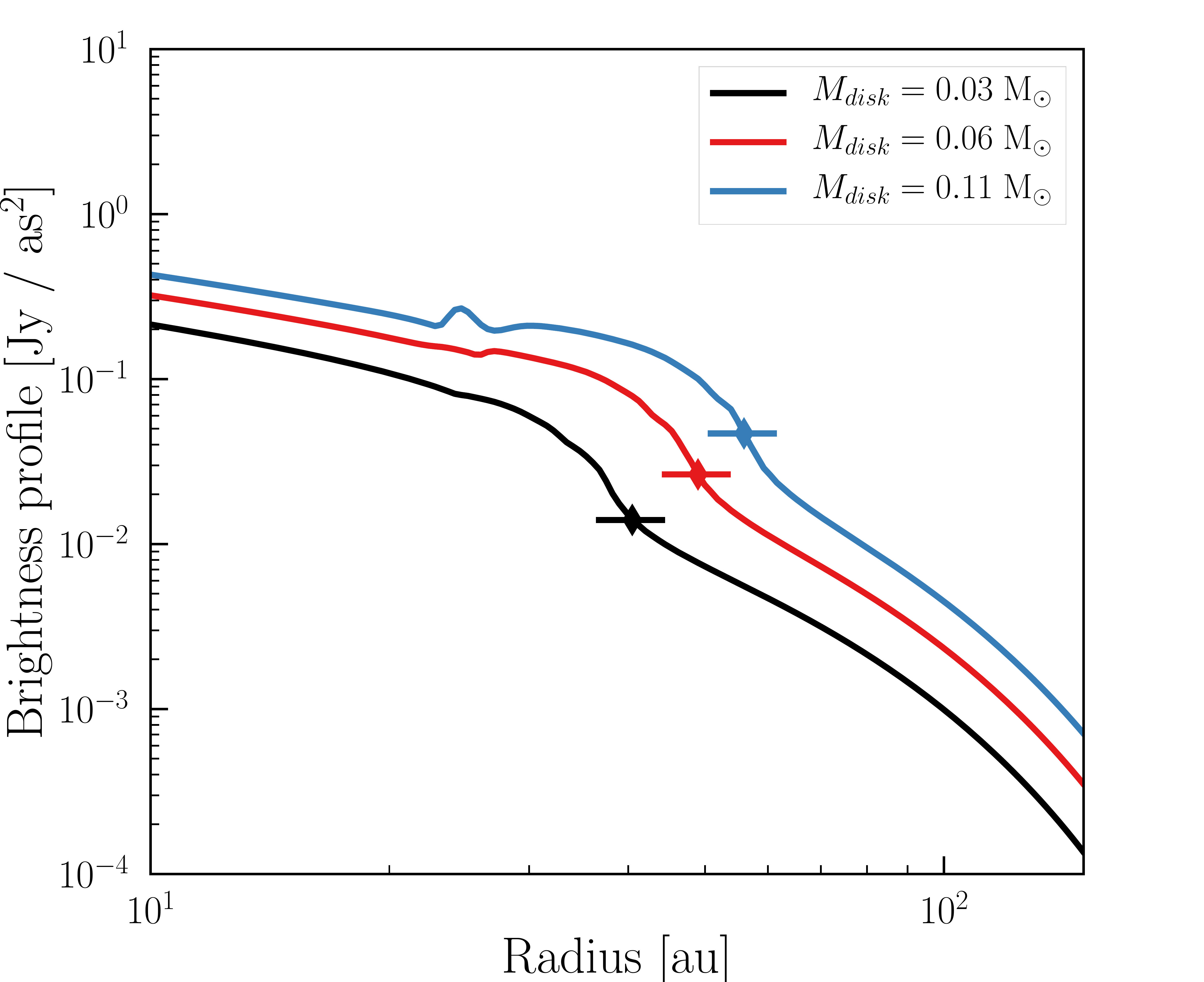}
    \caption{Dependence of the brightness profile on the disk mass. The dust line location, indicated by the dots on the profiles, shows a strong dependence on the disk mass.}
    \label{fig:disk_mass}
\end{figure}

The first requirement for any mass estimator is, by definition, to have a strong dependence on the disk mass, given in our case by Eq.\ref{eq:sigma_g}. To test this dependence we simulated the evolution of disks of different masses and study the changes in their brightness profiles, shown in Fig.\ref{fig:disk_mass}. The dust line locations in these profiles show a visible dependence on the disk mass. While this result confirms that the dust line location could be used as a good mass estimator, it does not not match the relation $M_{disk} \propto r^{-1}$ given by Eq.\ref{eq:sigma_g}. One possible explanation is that Eq.\ref{eq:sigma_g} only takes into account local dust evolution processes, neglecting the contribution to the dust distribution from particles drifting from outer radial locations. Our dust evolution model takes into account all these processes, possibly explaining the difference between our result and the prediction given by Eq.\ref{eq:sigma_g}. 

\subsection*{Stellar mass}
\begin{figure}
    \centering
    \includegraphics[width=8cm]{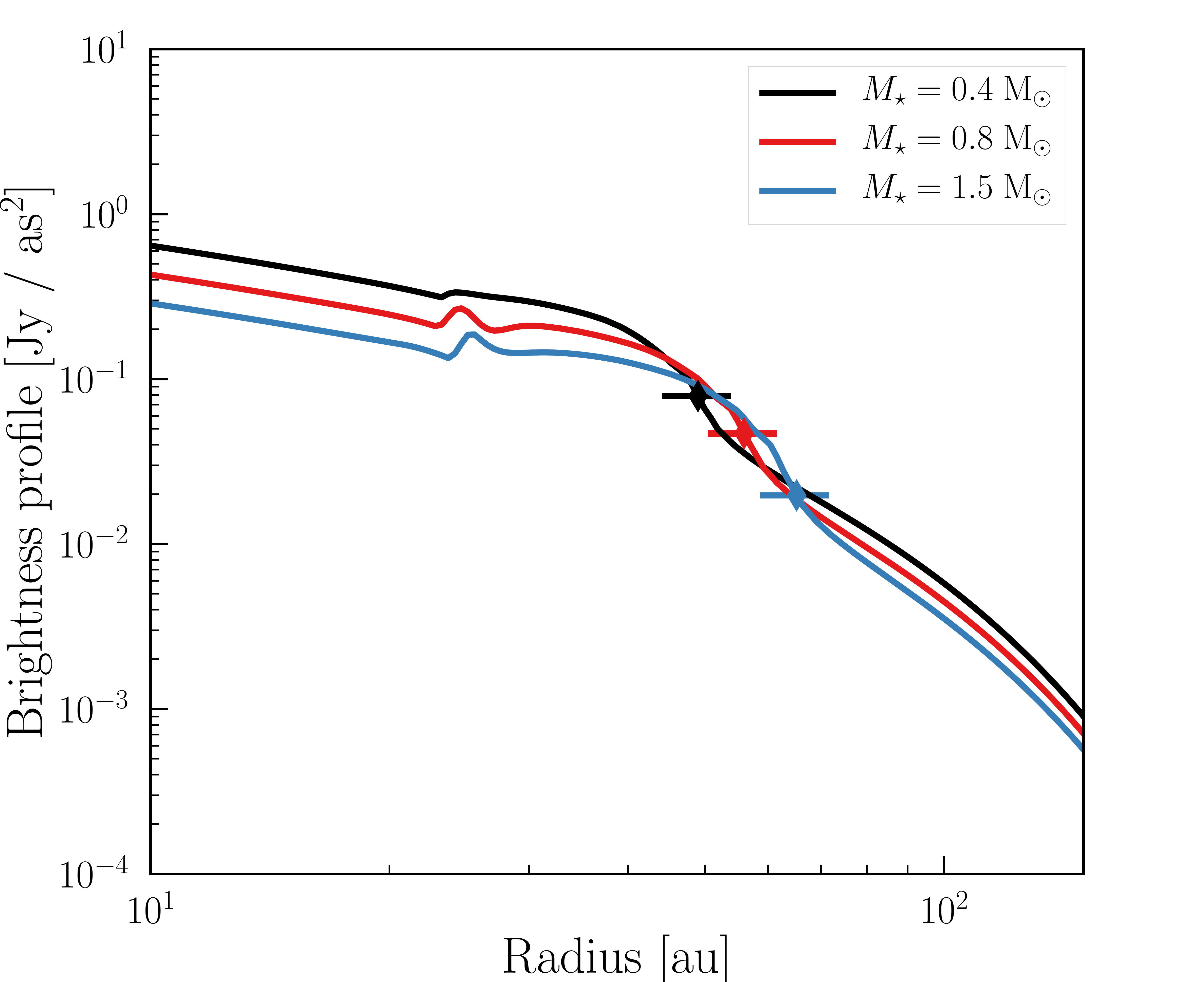}
    \caption{Dependence of the brightness profile on the stellar mass. The stellar mass affects the drift efficiency, however this parameter does not change significantly the dust line location.}
    \label{fig:stellar_mass}
\end{figure}

The amount of drift a grain undergoes depends on how well it is coupled to the gas structure. This is quantified by the Stokes number $\mathit{St} = t_{stop} \; \Omega_K$. Through the Keplerian frequency, the amount of drift depends on the stellar mass as $\mathit{St} \propto 1/\sqrt{M_\star}$ (e.g., \citealt{Birnstiel16, Chiang10}). This behavior is confirmed by the results presented in Fig.\ref{fig:stellar_mass}, where the dust line location depends on the stellar mass.\par
In more realistic cases $M_{disk}$ and $M_\star$ are correlated, and \cite{Pinilla20} suggest the disk mass to be a fraction of the stellar mass ($5\%$). Even in this case the dust line location is dependent on the $M_{disk} - M_\star$ parameter. Care has to be taken when applying the mass estimator to disks around stars of different masses.\par

\subsection*{Disk age}
\begin{figure}
    \centering
    \includegraphics[width=8cm]{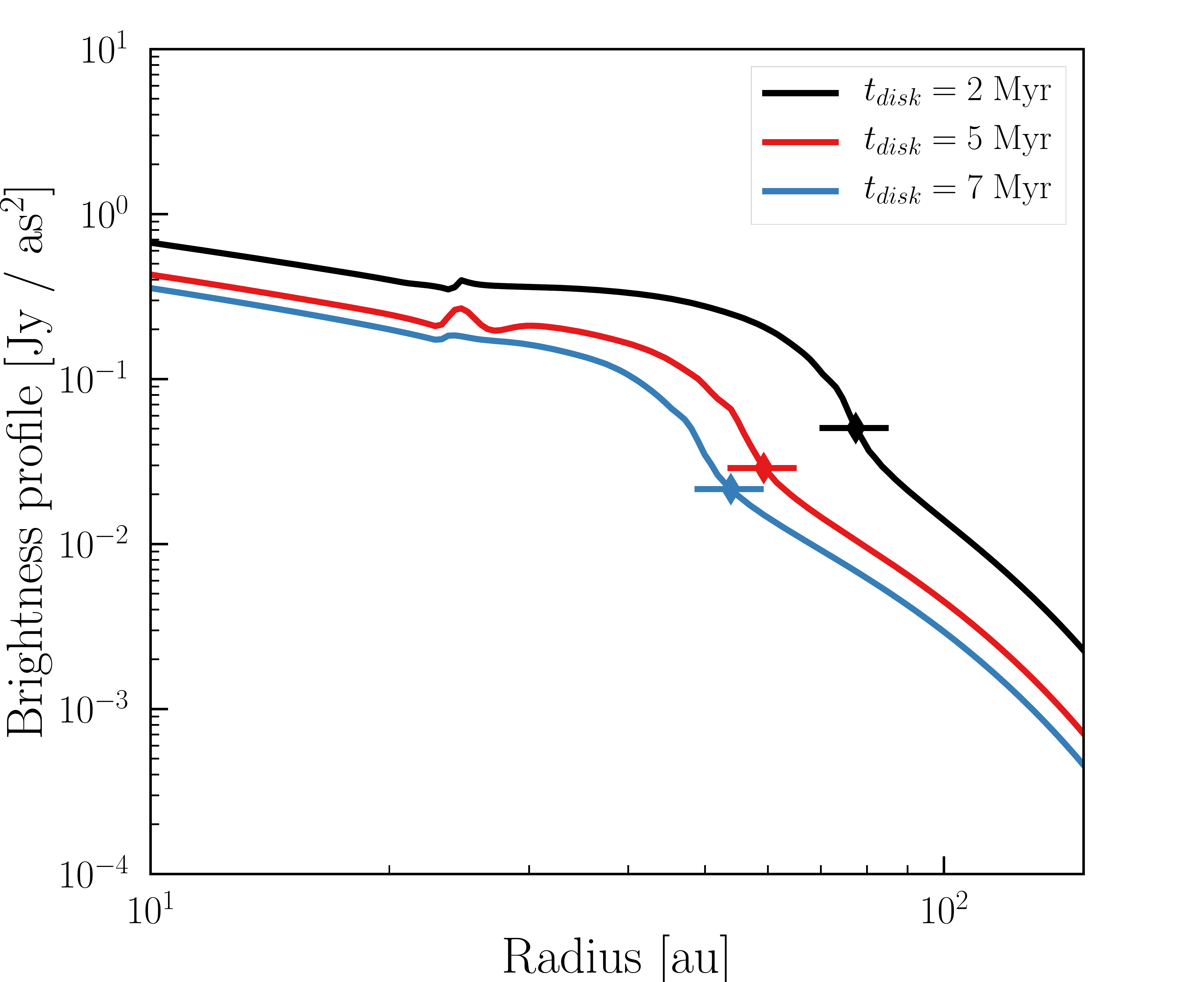}
    \caption{Dependence of the brightness profile on the age of the system. The location of the dust line has a strong dependence on this parameter, and any uncertainty on the disk age is directly propagated into the disk mass value.}
    \label{fig:age}
\end{figure}

The age of the disk determines how long particles have been drifting. Therefore, the dust line location is very sensitive to this parameter, as shown in both Eq.\ref{eq:sigma_g} and Fig.\ref{fig:age}. As the age of the disk increases, the grains drift over a larger distance toward the inner disk. This effect is degenerate with the disk mass, and it is not possible to distinguish a younger, less massive disk from an older and more massive one. Therefore, an independent age estimate is central to a reliable mass estimate. However, especially for young stars, the age of the system is subject to significant observational uncertainties. The proper tuning of this parameter is likely to be the major source of uncertainties of the model.

\begin{table*}
\caption{The dependence of the dust line location with the disk parameters, as described in Sec.\ref{Sec:params}.}         
\label{table:model_results}      
\centering                          
\begin{tabular}{c | c | c | c || c}
\hline\hline  
$M_{disk} \; [\mathrm{M_\sun}]$ & $M_\star \; [\mathrm{M_\sun}]$ & $t_{disk} \; [\mathrm{Myr}]$ & $\epsilon$ & Dust line location [au]\\
\hline                        
0.03, 0.06, 0.11 & 0.8 & 5 & 0.01 & 40, 49, 56\\
0.11 & 0.4, 0.8, 1.2 & 5 & 0.01 & 53, 56, 61\\
$0.05 \cdot M_\star$ & 0.4, 0.8, 1.2 & 5 & 0.01 & 30, 49, 69\\
0.11 & 0.8 & 2, 5, 7 & 0.01 & 72, 56, 52\\
0.11 & 0.8 & 5 & 0.01, 0.05, 0.1 & 56, 56, 56\\
\hline                                   
\end{tabular}
\end{table*}

\subsection*{Initial dust-to-gas ratio}
\begin{figure}
    \centering
    \includegraphics[width=8cm]{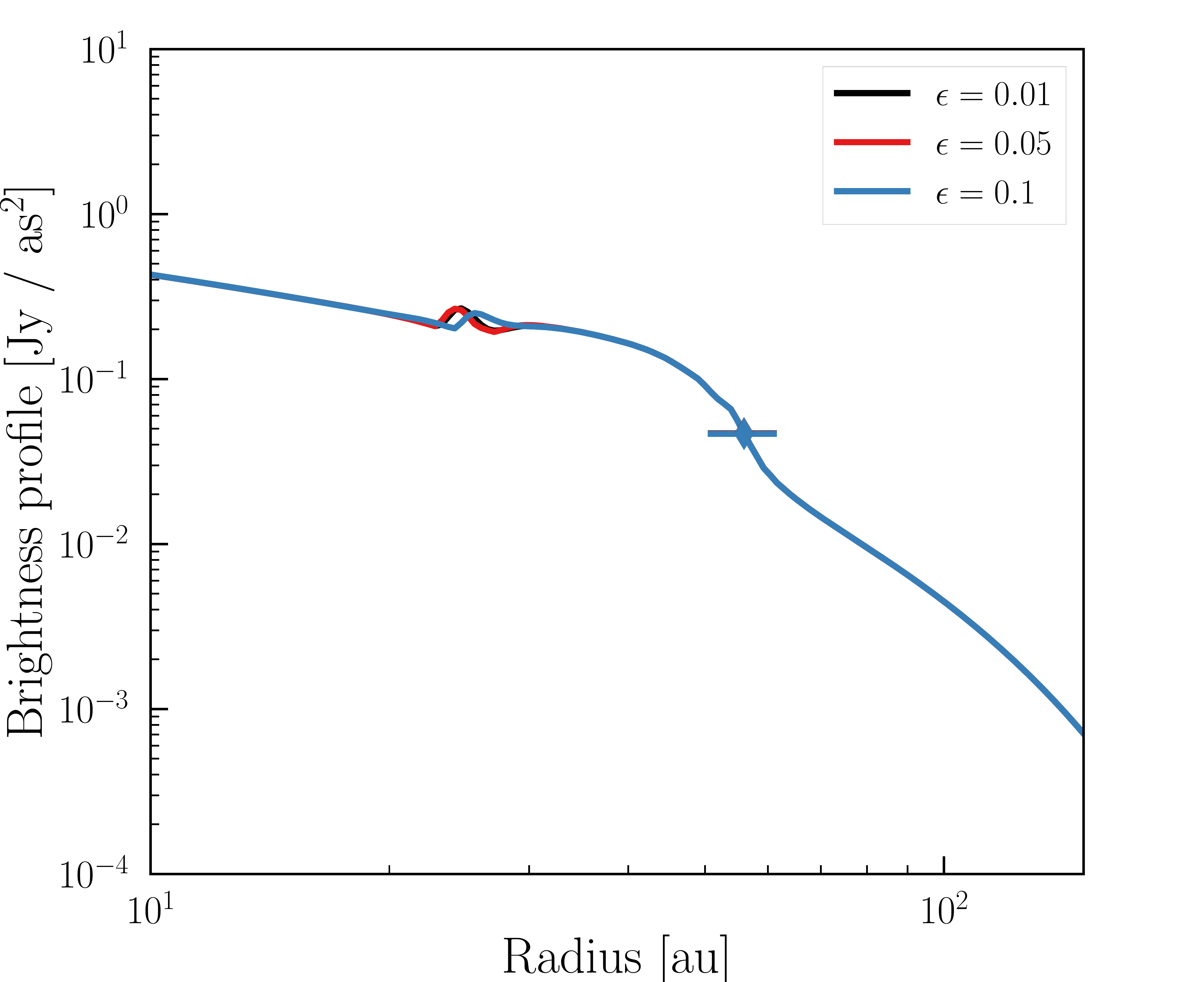}
    \caption{Dependence of the brightness profiles on the initial disk dust to gas ratio. The profiles are independent on the assumed initial ratio, suggesting that by this age most of the dust evolution already happened and it is not affecting our results.}
    \label{fig:d2g}
\end{figure}

The strength of this approach is that the mass estimate is independent of the dust-to-gas ratio, as we are looking at just the size of the emitting region and not its intensity profile. However Fig.\ref{fig:d2g} shows that not only the dust line location, but also the intensity does not depend on the dust-to-gas ratio assumed at the beginning of the disk evolution. The drift timescale \citep{Birnstiel12}, when equal to the disk age, can be written in terms of the dust-to-gas ratio:

\begin{equation}
    \label{eq:d2g_dustline}
    \epsilon = \frac{1}{t_{disk} \Omega_K}.
\end{equation}

\begin{figure}
    \centering
    \includegraphics[width=10cm]{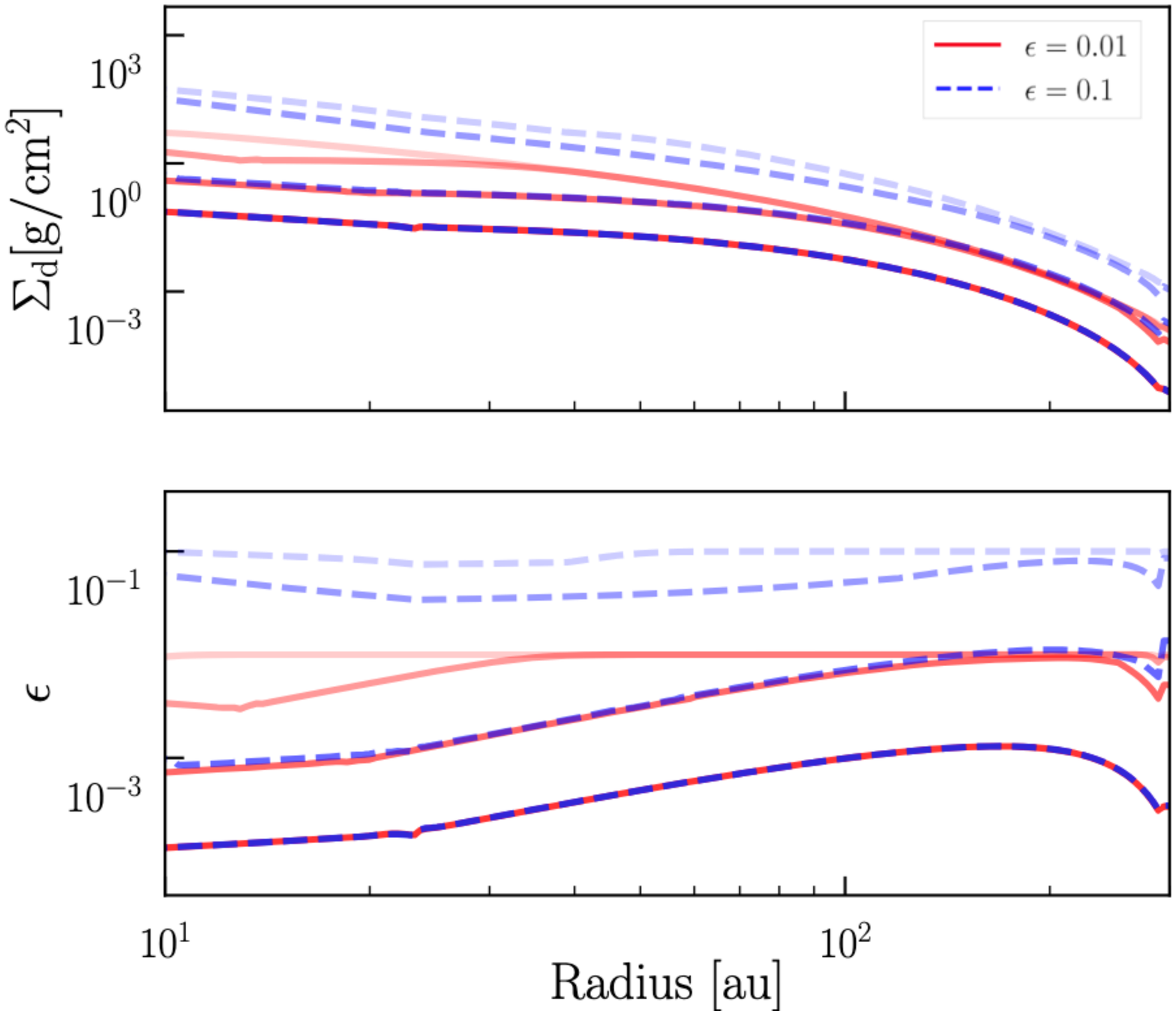}
    \caption{Time evolution, between 1 Myr and 5 Myr, of the dust surface density distribution (top) and the dust-to-gas ratio (bottom) assuming an initial dust-to-gas ratio of 0.01 (red) and 0.1 (blue), with darker lines for older ages. This plot shows how most of the dust evolution happens within the first few Myr. After that the dust density profile and the dust-to-gas ratio are independent of the assumed initial dust-to-gas ratio.}
    \label{fig:dust_evolution}
\end{figure}

Therefore, disks of different dust-to-gas ratios evolve toward this equilibrium value. Dust evolution happens mostly in the first few Myr, and within a lifetime of a Class II disk would reach the equilibrium value regardless how high or low it was initially. This is shown in Fig.\ref{fig:dust_evolution}, where we see the evolution of the dust surface density distribution and the dust-to-gas from 1~Myr to 5~Myr assuming an initial dust-to-gas ratio of 0.1 and 0.01. After a few Myr, the dust-to-gas ratio (and $\Sigma_d$ consequently) reach an equilibrium value dependent on the disk age, given by Eq.\ref{eq:d2g_dustline}, regardless of the assumed initial value.\par

%

\section{Population synthesis}
\label{sec:population}
\begin{figure}
    \centering
    \includegraphics[width=9cm]{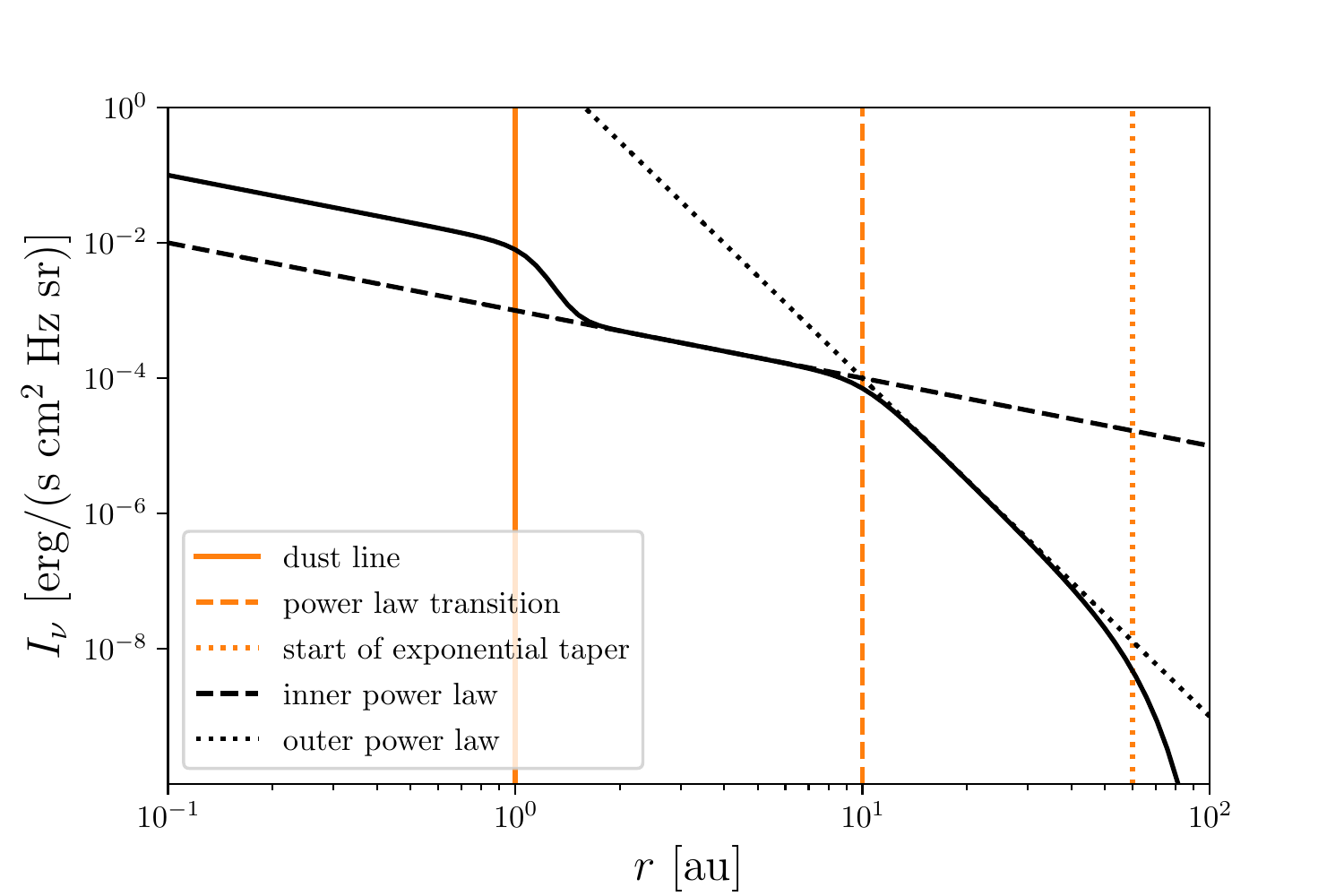}
    \caption{Function adopted to fit the simulated brightness profiles. This function accurately finds the outer edge of the dust emission.}
    \label{fig:dust_line_fit}
\end{figure}

In the previous section, we have shown how disk parameters affect the dust line location, suggesting that it is a good tracer for the disk mass. We now discuss how well Eq.\ref{eq:sigma_g} \citep{Powell17} estimates the mass of our synthetic disk models. We simulated a large number of disks on a grid of values for the parameters discussed in Sec.\ref{Sec:params}, summarized in Tab.\ref{table:population_params}. We then estimated the dust line location of four different dust populations from the 0.087 cm, 0.1 cm, 0.3 cm, and 0.9 cm emission profiles. In Sec.\ref{Sec:dustline}, we showed that the location enclosing 99\% of the dust emission at a wavelength $\lambda$ is well representing the dust line location of grains of size $\lambda / 2 \pi$. Alternatively, to analyze more consistently the large number of disks of this section, we took a more error-proof approach by fitting the dust emission profiles with a  fitting function such as the one shown in Fig.\ref{fig:dust_line_fit}. This function is composed by an inner and outer power law (inner and outer disk)  with an exponential taper, plus a sigmoid function at the dust line location to reproduce the characteristic steep emission drop.  

This approach assumes a regular, monotonically decreasing profile, a condition satisfied in our synthetic population, but not in more realistic disks with evidence of substructures such as rings and gaps. In this section, we estimate the dust line location by fitting the synthetic profiles to have a more error-proof estimate, as necessary when working with a large number of disks. In the other sections of the paper, we again place the dust line on the location enclosing 99\% of the dust emission, as this definition works also in the presence of radial features and can be easily double-checked when working with a small number of disks.\par

\begin{figure}
    \centering
    \includegraphics[width=9cm]{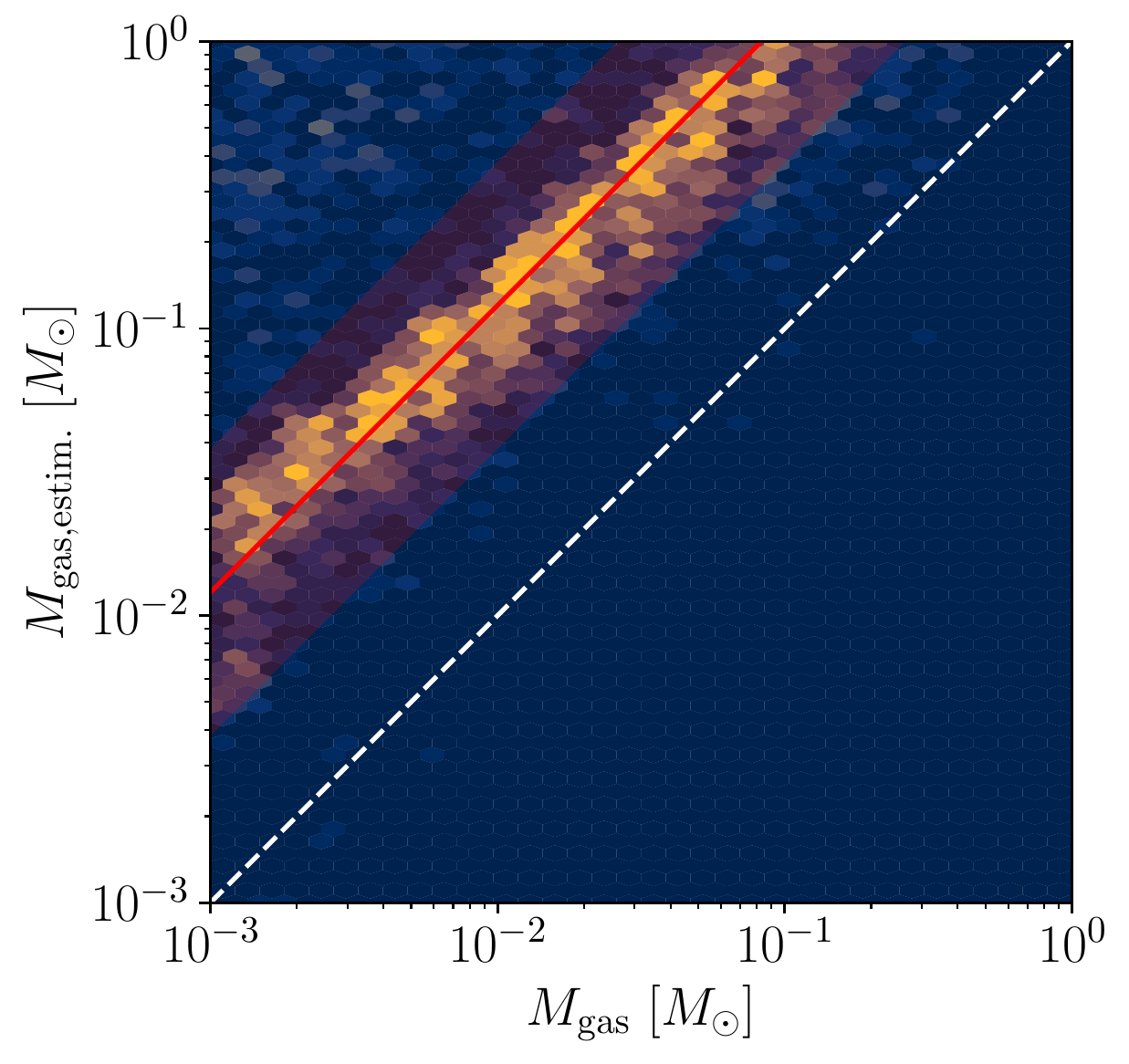}
    \caption{Comparison between input disk masses and the estimates from the dust line location. The shaded red area includes the best $68\%$ of the disks, the red line is the median mass of these disks, and the white line is the correlation function for a perfect match. The dust line location systematically overestimates the disk mass by about one order of magnitude.}
    \label{fig:population}
\end{figure}

Using these dust line location estimates, we followed the same scheme used in \cite{Powell17}. We calculated the gas surface density at the dust line locations using Eq.\ref{eq:sigma_g}, which we then fitted to a Lynden-Bell-Pringle self-similar profile, as in Eq.\ref{eq:LBP}, to estimate the gas distribution and its total mass. We compared these mass estimates to the input model masses to test how tight the correlation between the estimated mass and the model mass is. The result of this study, shown in Fig.\ref{fig:population}, is in agreement with the results in Sec.\ref{Sec:params}, supporting the strong dependence of the dust line location on the disk mass, but also provide further interesting information.\par

\begin{table}
\caption{Input parameters of our synthetic disk population.}             
\label{table:population_params}      
\centering                          
\begin{tabular}{c c}        
\hline\hline                 
Parameter & Value \\    
\hline                        
   $M_{disk}$ & $[10^{-3} - 0.2]$ M$_\sun$\\
   $M_*$ & [0.2 - 2] M$_\sun$\\
   $\alpha$ & $[10^{-4} - 0.03]$\\
   $v_{frag}$ & [1 - 20] m/s\\
   $r_c$ & [30 - 200] au\\
   $t_{disk}$ & 5 Myr\\
\hline                                   
\end{tabular}
\end{table}

The first interesting point is how well correlated the model disk masses and the dust line based estimates are, although with some scattering due to the dependence on the other disk parameters as we discussed already in Sec.\ref{Sec:params}. However, the plot also shows that mass estimates derived from Eq.\ref{eq:sigma_g} do overestimate systematically the disk mass. Indeed, Eq.\ref{eq:sigma_g} is derived from local considerations on the timescales of the processes involved in dust evolution at the dust line location, as we discussed in Sec.\ref{Sec:model}. However, the dust line locations of different grain populations are not independent, as the total dust mass flux has to be conserved as the grains drift toward the inner disk. When assuming that the dust evolution happens on a timescale of about the age of the disk, we are implicitly assuming that the dust mass flowing through the dust line is negligible compared to its total mass at this location. This is a good approximation in the outer disk, the main mass reservoir, but it breaks down in the inner disk. As a matter of fact, the gas surface density estimates from Eq.\ref{eq:sigma_g} for our synthetic population get worse at smaller radial locations. Since Eq.\ref{eq:sigma_g} holds in the outer disk, it provides sufficient constraints to the disk mass from the dust line of small grains, explaining the strong correlation seen in Fig.\ref{fig:population}.\par

\begin{figure}
    \centering
    \includegraphics[width=9cm]{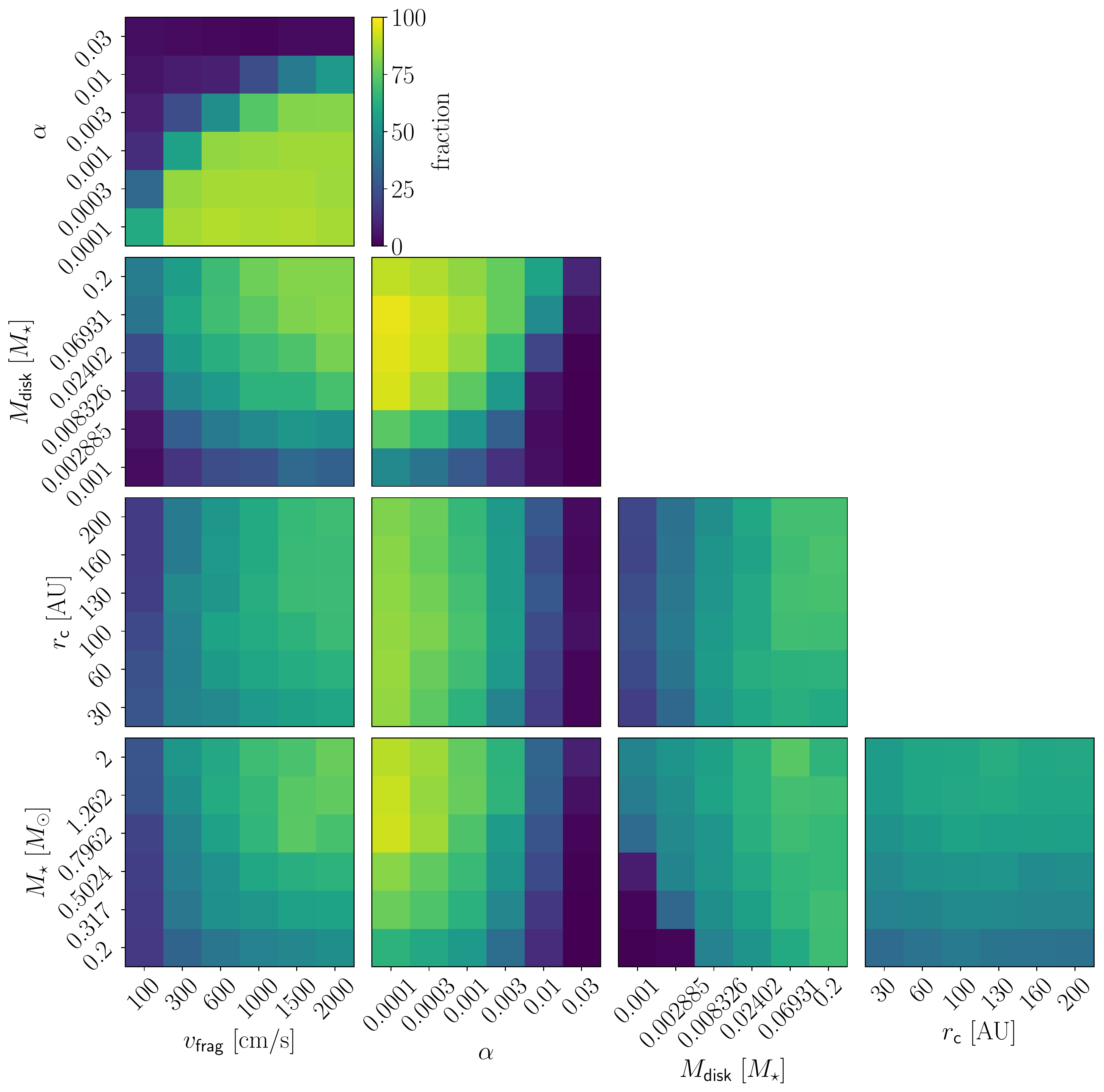}
    \caption{Projection of the posterior probability distribution of our fit. The disks are evenly distributed over the parameter space, except for very low disk mass or high turbulence.}
    \label{fig:corner}
\end{figure}

Other interesting information can be derived by fitting a power law to the data. From the posterior probability distribution function of the disk parameters, we can check for which parameter range the dust line location correctly predicts the disk mass. Fig.\ref{fig:corner} shows the parameter distribution for the disks whose mass estimate is within a factor of 3 of the calibrated relation. This plot states what fraction of these simulations with the given parameter choices are within this constraint. The plot shows that the correlation between the dust line mass estimate and the model masses breaks down for highly turbulent disks ($\alpha \gtrsim 10^{-2}$) and for very low disk masses ($\lesssim 10^{-4}$ M$_\sun$). In these cases, the assumption of drift dominated dust evolution is no longer satisfied, and the dust line location is less and less dependent on the disk mass. However, these values are unrealistic in most observed disks, and therefore this does not affect the reliability of mass estimates based on the dust line location. 

\section{Surface density distributions with pressure bumps}
\label{sec:structures}
We have shown that the dust line location is a reliable tracer for the disk mass in the case of a smooth gas surface density distribution, confirming the general concept of \cite{Powell17, Powell19}. We stress here that the results of the population synthesis study given in the previous section do not affect our method: by simulating the dust evolution, rather than relying on a calibrated Powell expression, our results are independent on the assumptions from which Eq.\ref{eq:sigma_g} is derived. Moreover, disks often show substructures, such as rings and gaps in their brightness distribution, often interpreted as regions of high/low pressure, or pressure bumps \citep[e.g.,][]{Pinilla12, Rosotti20}. Because pressure bumps help to reduce or completely suppress the radial drift, the outer edge of the dusty disk or dust line is directly influenced by the potential presence of pressure bumps. To include density bumps in our gas distribution we assumed Gaussian perturbations as in \cite{Pinilla20}:

\begin{equation}
    \Sigma'(r) = \Sigma(r) \cdot (1 + B(r)),
\end{equation}

\noindent with    

\begin{equation}
    B(r) = A\; \exp{- \left( \frac{(r-r_p)^2}{2w^2} \right)},
\end{equation}

where $\Sigma'(r)$ is the unperturbed density from Eq.\ref{eq:LBP} and $A$ is the amplitude, $r_p$ the center, and $w$ the width of the Gaussian perturbation, that we assumed to be equal to the pressure scale-height $H_p$. The width has to be larger or equal to $H_p$ to ensure the stability of the bumps (e.g., \citealt{Pinilla12, Dullemond18}). A narrower bump has a higher pressure gradient and a stronger effect on the dust evolution. This is studied by setting $w = H_p$. We considered two kinds of bumps: weak bumps with $A = 1$, and strong bumps with $A = 4$. Hydrodynamical simulations from \cite{Zhang18} show that bumps with $A = 1$ and $A = 4$ resemble the presence of a giant planet of mass 0.3 $\mathrm{M_{Jup}}$ and 1 $\mathrm{M_{Jup}}$, respectively. For this range of masses, the bump width does not appear to change with $A$, justifying our assumption to use the same width. \par

As discussed in Sec.\ref{Sec:model}, particle trapping is likely to happen inside radial substructures. Since dust lines differ in their location at different wavelengths, this suggests a different distribution for particles of different sizes. If a ring is an efficient particle trap, we would expect an increase in the dust continuum emission at every wavelength at the trap location. However the dust evolution is still drift dominated outside of the density bump, and dust lines at a different location than the bump radius could be used to estimate the disk mass. When a density bump affects the dust line, such an effect could be accounted for by dust evolution models to yield a correct mass estimate.\par

\begin{figure}
    \centering
    \includegraphics[width=8cm]{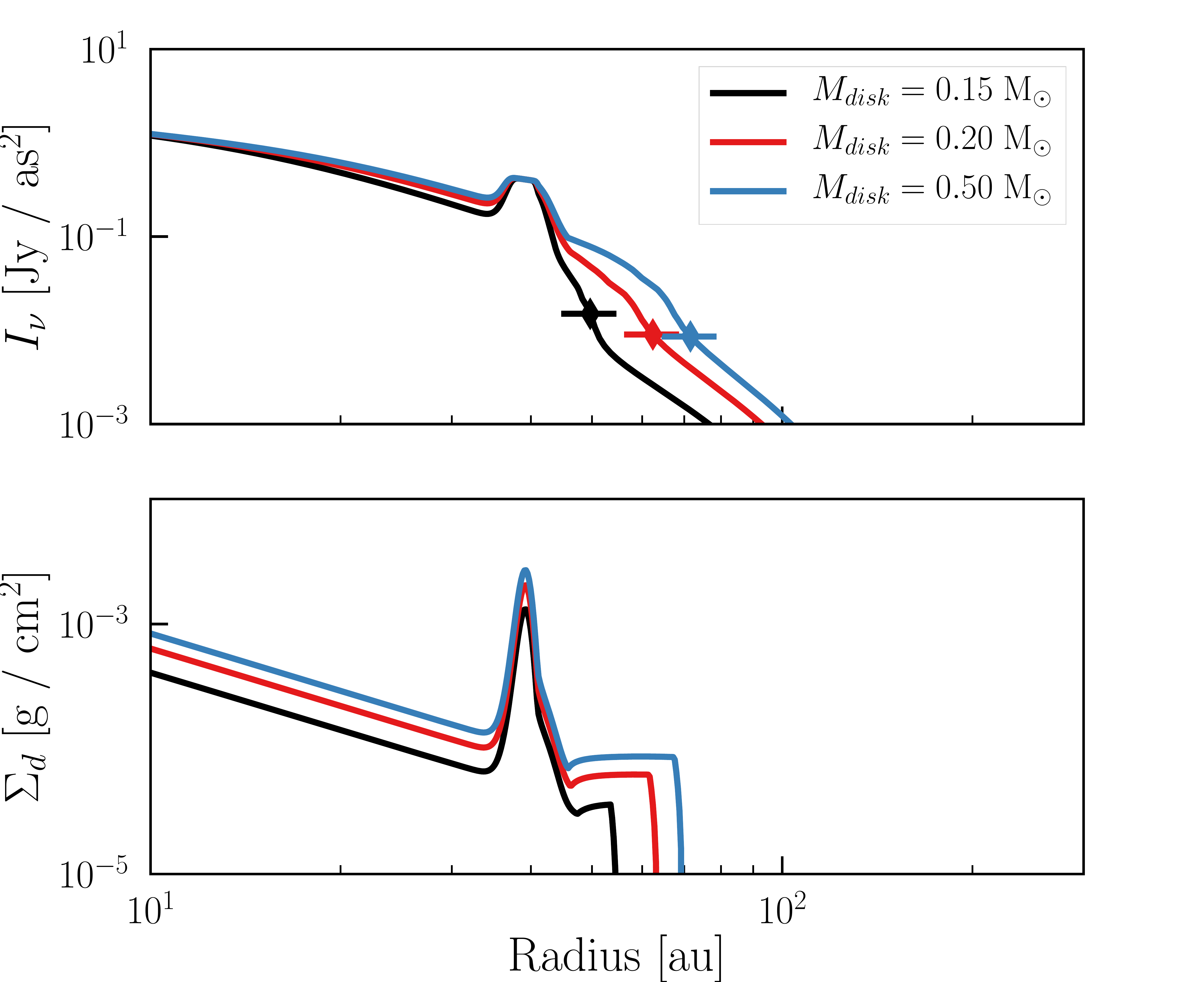}
    \caption{Dependence of the dust line location on the disk mass in the presence of a weak bump in the gas density at 40 au. The upper plot shows the dust emission profile at $\lambda = 0.87$ mm, as predicted by the model. The dust line location, as defined in the text, is highlighted on the profiles. The lower plot shows the surface density distribution of the most emitting grains. The dust line location in the first plot matches the outer edge of the density distribution in the lower plot for the same disk mass.}
    \label{Fig:bump_40au}
\end{figure}

\begin{figure}
    \centering
    \includegraphics[width=8cm]{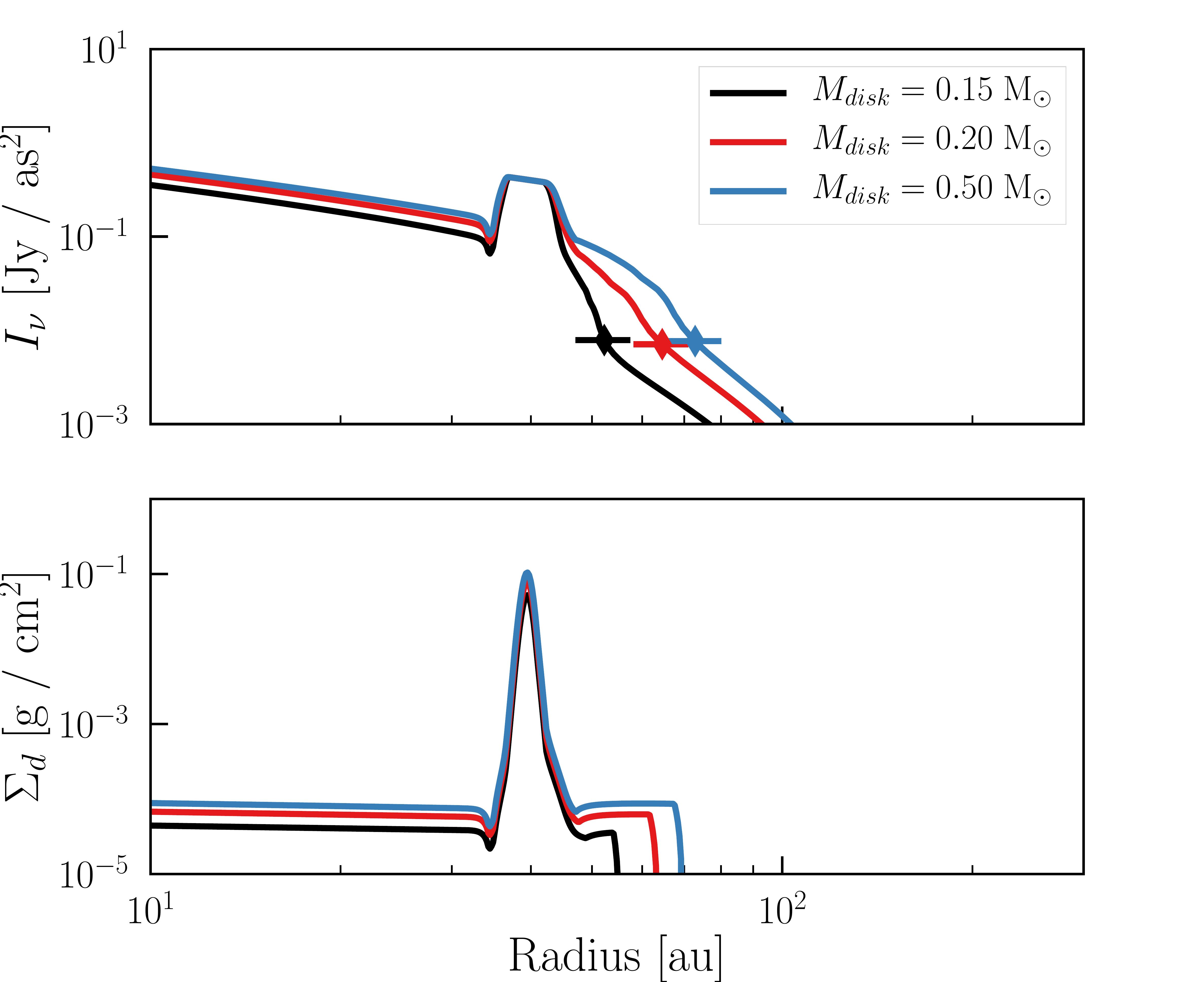}
    \caption{Dependence of the dust line location on the disk mass in the presence of a strong bump in the gas density at 40 au. The upper plot shows the dust emission profile at $\lambda = 0.87$ mm, as predicted by the model. The dust line location, as defined in the text, is highlighted on the profiles. The lower plot shows the surface density distribution of the most emitting grains. The dust line location in the first plot matches the outer edge of the density distribution in the lower plot for the same disk mass.}
    \label{Fig:40au_trapping}
\end{figure}

The first test is to study how a bump affects the dependence of the dust line on the disk mass by evolving several disks with different masses and adding a bump at 40 au. The used disk parameters are shown in Tab.\ref{table:disk_parameters}, and the simulation results in Fig.\ref{Fig:bump_40au} and Fig.\ref{Fig:40au_trapping} with a narrow and wide bump ($A = 1$ and $A = 4$, respectively). The size of the emitting region is still tracing the dust line location, which is still strongly dependent on the disk mass. The location of the dust line depends on the bump amplitude A, but it is still mainly determined by the disk mass.\par

Another parameter that can affect the dust line location is the bump location. We expect bumps in the outer disk to have a stronger effect on the grain evolution, as more grains will cross the bump during the disk lifetime. Our models show that bumps outside the dust line have a larger effect on the dust line location than bumps inside the dust line location, however this does not affect the applicability of the method. 

\begin{table}
\caption{Disk mass $M_{disk}$ and critical radius $r_c$ (from Eq.\ref{eq:LBP}), initial dust-to-gas ratio $\epsilon_0$, stellar mass $M_\star$ and luminosity $L_{star}$, and age of the system $t_{disk}$ used in the dust evolution model, as measured in TW Hya as a test case.}             
\label{table:disk_parameters}      
\centering                          
\begin{tabular}{c c}        
\hline\hline                 
Parameter & Value \\    
\hline                        
   $M_{disk}$ & $[0.01 - 0.2]$ M$_\sun$ \\
   $r_c$ & $30$ au\\
   $\epsilon_0$ & $0.01$\\   
   $M_*$ & 0.8 M$_\sun$\\
   $L_*$ & 0.28 L$_\sun$\\
   $t_{disk}$ & 5 Myr\\
   $A$ & [1, 4]\\
   $r_p$ & 40 au\\
   $w$ & 4.3 au\\
\hline                                   
\end{tabular}
\end{table}

\begin{figure}
    \centering
    \includegraphics[width=8cm]{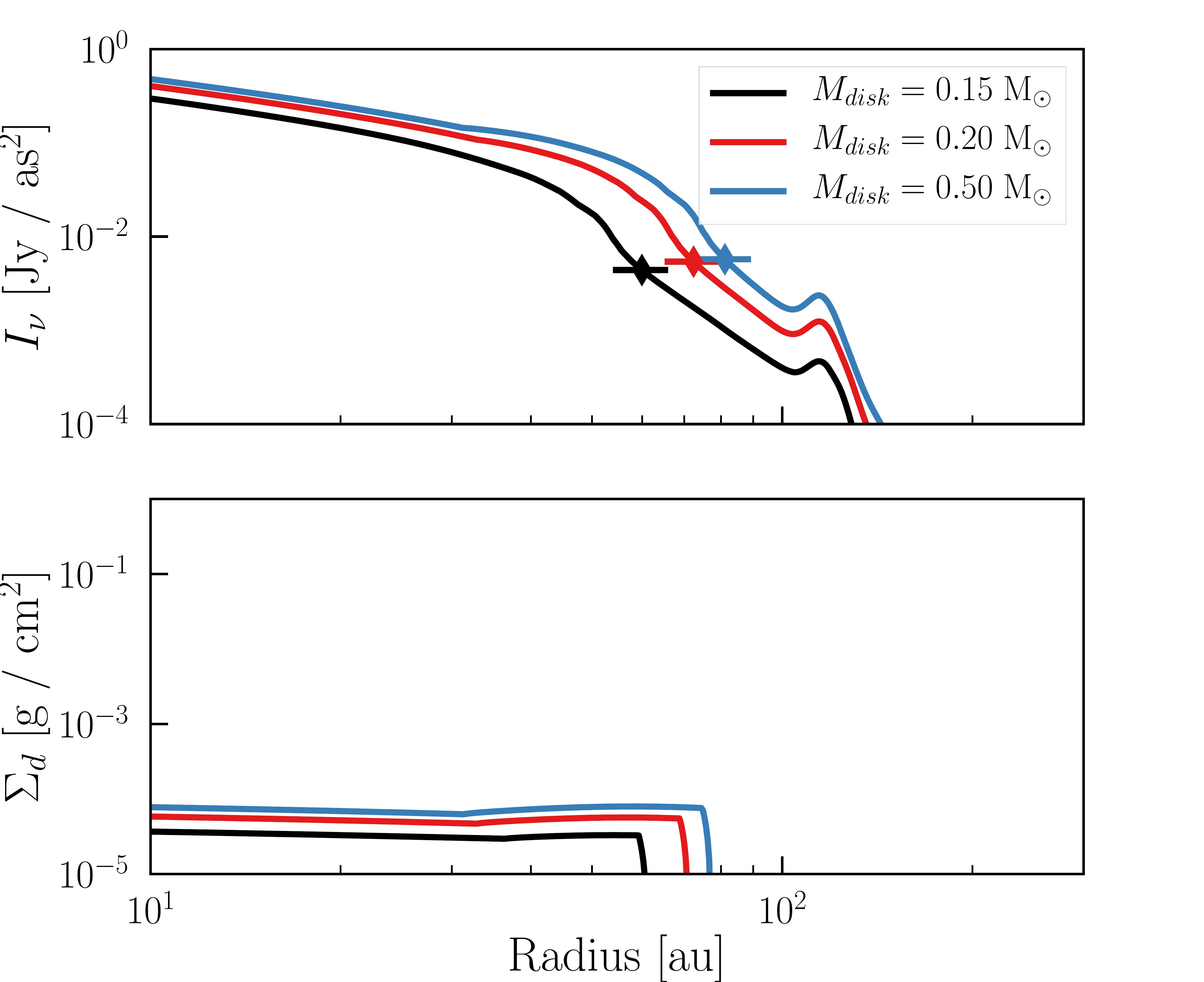}
    \caption{Dependence of the dust line location on the disk mass in the presence of a weak bump in the gas density at 120 au. The upper plot shows the dust emission profile at $\lambda = 0.87$ mm, as predicted by the model. The dust line location, as defined in the text, is highlighted on the profiles. The lower plot shows the surface density distribution of the most emitting grains. The dust line location in the first plot matches the outer edge of the density distribution in the lower plot for the same disk mass.}
    \label{Fig:bump_120au}
\end{figure}

\begin{figure}
    \centering
    \includegraphics[width=8cm]{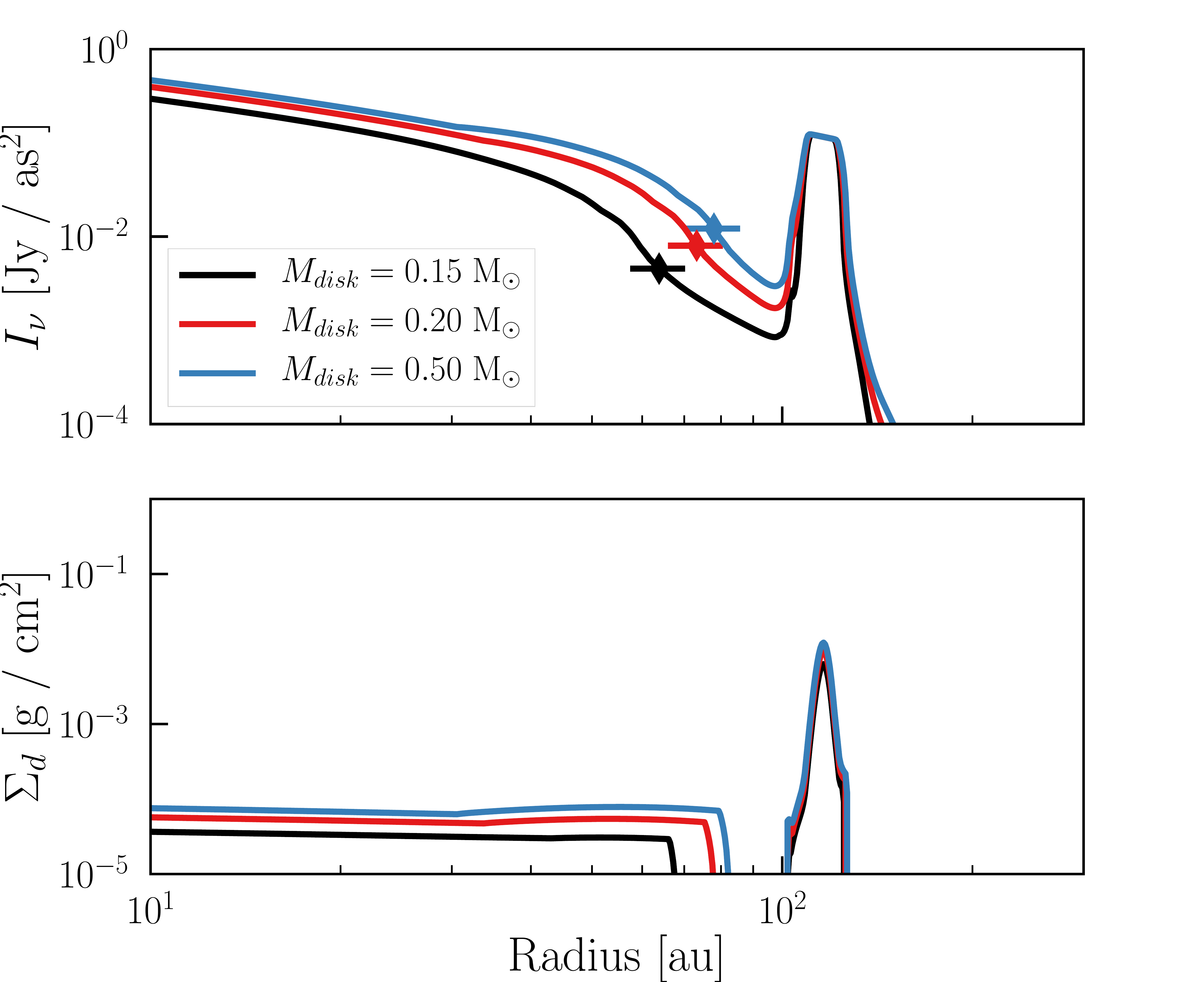}
    \caption{Dependence of the dust line location on the disk mass in the presence of a strong bump in the gas density at 120 au. The upper plot shows the dust emission profile at $\lambda = 0.87$ mm, as predicted by the model. The dust line location, as defined in the text, is highlighted on the profiles. The lower plot shows the surface density distribution of the most emitting grains. The dust line location in the first plot matches the outer edge of the density distribution in the lower plot for the same disk mass.}
    \label{Fig:120au_trapping}
\end{figure}

Such a case is shown in Fig.\ref{Fig:bump_120au} for a disk model with an outer bump with an amplitude $A = 1$. The profiles show a drop in the brightness typical of the dust line inside the bump, whose location can be blindly used to provide a disk mass estimate. This result does not change with the bump amplitude as shown in Fig.\ref{Fig:120au_trapping} for a disk model with a bump of amplitude $A = 4$. In this disk model, more grains gather at the bump, but in the region inside the trap we can still observe a drift-defined dust line. As in the case of a bump in the inner disk, the dust line location is dependent on the bump amplitude, but it is still drift dominated and it can be used to provide a mass estimate.\par

When the disk presents a more complex structure featuring multiple bumps, the dust line location may lose its dependence on the disk mass. As the number of radial substructures increases, the more extended are the regions of the disk that are not drift dominated, and dust lines in these regions will have weak dependence on the disk mass. In this case, resolved observations at different millimeter wavelengths should all show similar brightness profiles. In this complex case other methods need to be used to estimate the disk mass.\par

\begin{figure}
    \centering
    \includegraphics[width=8cm]{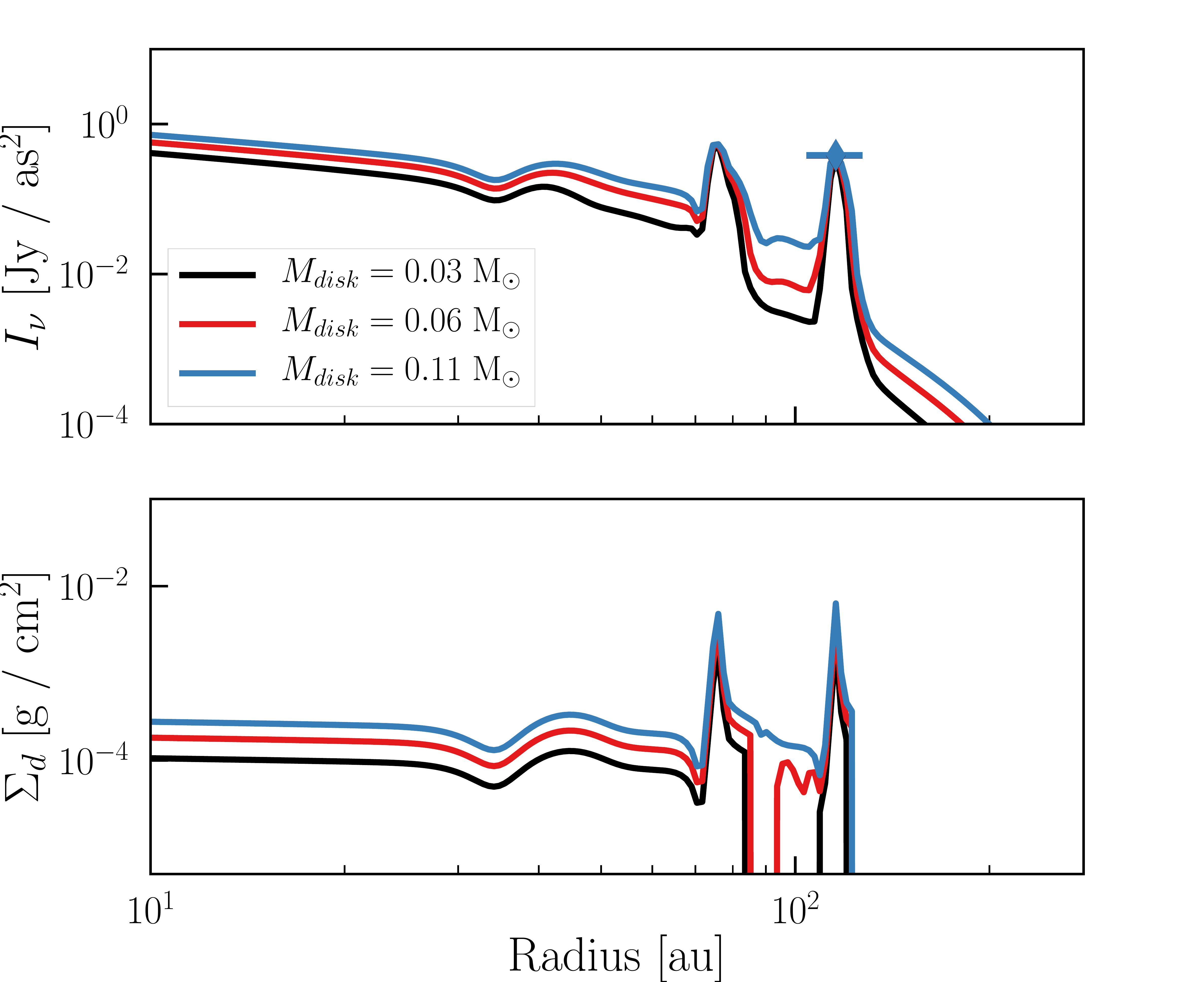}
    \caption{Dependence of the dust line location on the disk mass in the presence of weak bumps at 40 au, 80 au and 120 au. The upper plot shows the dust emission profile at $\lambda = 0.87$ mm, as predicted by the model. The dust line location, as defined in the text, is highlighted on the profiles. The lower plot shows the surface density distribution of the most emitting grains. Both the dust line location and the outer edge of the dust surface density distribution trace the location of the outer bump and do not depend on the disk mass.}
    \label{Fig:multiple_bumps}
\end{figure}

In general, modeling the dust evolution is necessary to provide a reliable mass estimate. In Fig.\ref{Fig:multiple_bumps} we show an example of a disk with multiple radial features where the dust line location is independent on the disk mass, and is tracing the location of a pressure bump. This disk model has the same parameters as in Tab.\ref{table:disk_parameters}, with weak bumps at 40 au, 80 au and 120 au.\par 

\section{Applicability to individual disks}
\label{Sec:disks}
In this section we show results from disk evolution models of TW~Hya, with smooth structures, and AS~209, with very sharp and strong structures. We modeled the dust evolution of these two disks to explore the relation between dust line positions, bumps and disk masses. We then compared the disk masses inferred from the dust line locations with the input model parameters.\par

In an old disk with a smooth structure, such as TW~Hya, we would expect the dust evolution to be drift dominated. Bumps in the gas density profile could cause a departure from the simple relation between local surface density and dust line location given by Eq. \ref{eq:sigma_g}. However, the dust line location should still maintain a strong correlation with the disk mass, and a mass estimate could still be provided by matching the model prediction of dust line locations with dust continuum data.\par

When dealing with a structure with a higher contrast between gaps and rings, such as in AS~209, the global effect of drift on dust evolution is much weaker. Stronger radial features make the correlation between disk mass and dust lines weaker, and even if the grains are not trapped by the substructures, there will be a large uncertainty in the mass estimate.\par

The age of the disk is typically assumed to be the same as that of the host star which are subject to significant uncertainties. Given the form of Eq.\ref{eq:sigma_g}, the disk mass scales linearly with the age estimate. An older disk implies that grains are allowed to drift to more inward radial locations, therefore an increase in the age is indistinguishable from a decrease in the disk mass in determining the dust line location. The disk age is likely the highest source of uncertainty in the model, as shown in Sec.\ref{Sec:params}.

\subsection{TW Hya}
TW~Hya is one of the most well-studied disks, and it is a good benchmark to test our model. In the literature the TW Hya disk mass has been estimated using integrated dust emission \citep{Andrews12, Menu14}, CO line emission \citep{Rosenfeld12, Kama16} and HD line emission \citep{Bergin13, McClure16, Trapman17}. The mass of TW~Hya was also derived by \cite{Powell17} using the dust line location as we described. All these estimates are reported in Tab.\ref{table:TWHya_mass_estimates}.\par

The discrepancy between the mass derived from the dust line can be explained by a few factors. Evolving disks can be depleted of dust \cite[e.g.,][]{Birnstiel12}, and optical depth effect can lead to a mass underestimation when using dust emission. HD-derived mass is very sensitive to the assumed thermal structure: \cite{Trapman17} suggest a disk mass for TW~Hya about one order of magnitude lower than \cite{Bergin13} using a different disk structure. With regard of CO line emission, there is evidence of severe carbon depletion from the gas phase \citep{Kama16, Miotello17} that question the reliability of CO-based estimates. A survey of the Lupus star-forming region by \cite{Ansdell16} showed that CO-derived disk masses are not compatible with the measured accretion rates. An independent mass estimator, such as the dust line location, can lead to a better understanding of the disk structure from the discrepancy between mass estimates from different techniques.\par

\begin{table}
\caption[]{\label{nearbylistaa2} TW Hya mass estimates from different methods.}             
\label{table:TWHya_mass_estimates}      
\centering                          
\begin{tabular}{c | c}        
\hline\hline                 
Integrated dust emission & 0.018 M$_\sun$ \footnote{}\\
CO Line Emission & 0.003 M$_\sun$\footnote{}\\
HD Line Emission & > 0.05 M$_\sun$\footnote{}\\
    & $[7.7 \cdot 10^{-3}, \; 2.3 \cdot 10^{-2}] \; \mathrm{M_\sun}$\footnote{}\\
Dust Line Location & 0.11 M$_\sun$\footnote{}\\    
\hline                             
\end{tabular}
\tablebib{(1)~\citet{Andrews12, Menu14};
(2) \citet{Rosenfeld12, Kama16}; (3) \citet{Bergin13}; (4) \citet{Trapman17}, (5) \citet{Powell17}.
}
\end{table}

For our representative dust evolution model of TW~Hya we used the same parameter setup as in \cite{Powell17} and checked how well we could reproduce the dust continuum observations. Spatially resolved CO observations \citep{Rosenfeld12} are well reproduced by the Linden-Bell \& Pringle solution (Eq.\ref{eq:LBP}) with best fit parameters $r_c = 30 \; au$ and $\gamma = 1$. As $\Sigma_c$ depends on the assumed CO-to-H$_2$ ratio, this remains a free parameter. The C$^{18}$O column density profile derived in \cite{Zhang17} features a peak in the gas surface density at 70 au. We included this feature by adding a scaling factor to our gas density profile (see also \citealt{Huang18}):

\begin{equation}
    \Sigma(r) = \Sigma_c \left(  \frac{r}{r_c} \right)^{-\gamma} \exp \left[ -\left(  \frac{r}{r_c} \right)^{2-\gamma} \right] \times f(r),
\end{equation}

where f is setting the shape of the bump in the gas column density profile:

\begin{equation}
    f(r) = 
    \begin{cases} 
    1 + A \exp{ \left( -\frac{(r - R_{bump})^2}{2 \sigma_{in}} \right)}, & r < R_{bump}\\
    1 + A \exp{ \left( -\frac{(r - R_{bump})^2}{2 \sigma_{out}} \right)}, & r > R_{bump},\\
    \end{cases}
\end{equation}

with the model parameters are as listed in Tab.\ref{table:TWHya_params}.

\begin{table}[H]
\caption{Parameter values assumed for TW Hya.}             
\label{table:TWHya_params}      
\centering                          
\begin{tabular}{c c | cc}        
\hline\hline                 
Parameter & Value & Parameter & Value\\    
\hline                        
   $M_*$ & 0.8 M$_\sun$ & $\sigma_{in}$ & 12 au\\
   $M_{disk}$ & 0.11 M$_\sun$ & $\sigma_{out}$ & 6 au\\
   $r_c$ & 30 au & $t_{disk}$ & 5 Myr\\
   $\gamma$ & 0.9 & $T_{10}$ & 40 K\\
   A & 3 & $a_{turb}$ & 0.01\\
   $R_{bump}$ & 70 au\\
\hline                                   
\end{tabular}
\end{table}

\begin{figure}[H]
    \centering
    \includegraphics[width=8cm]{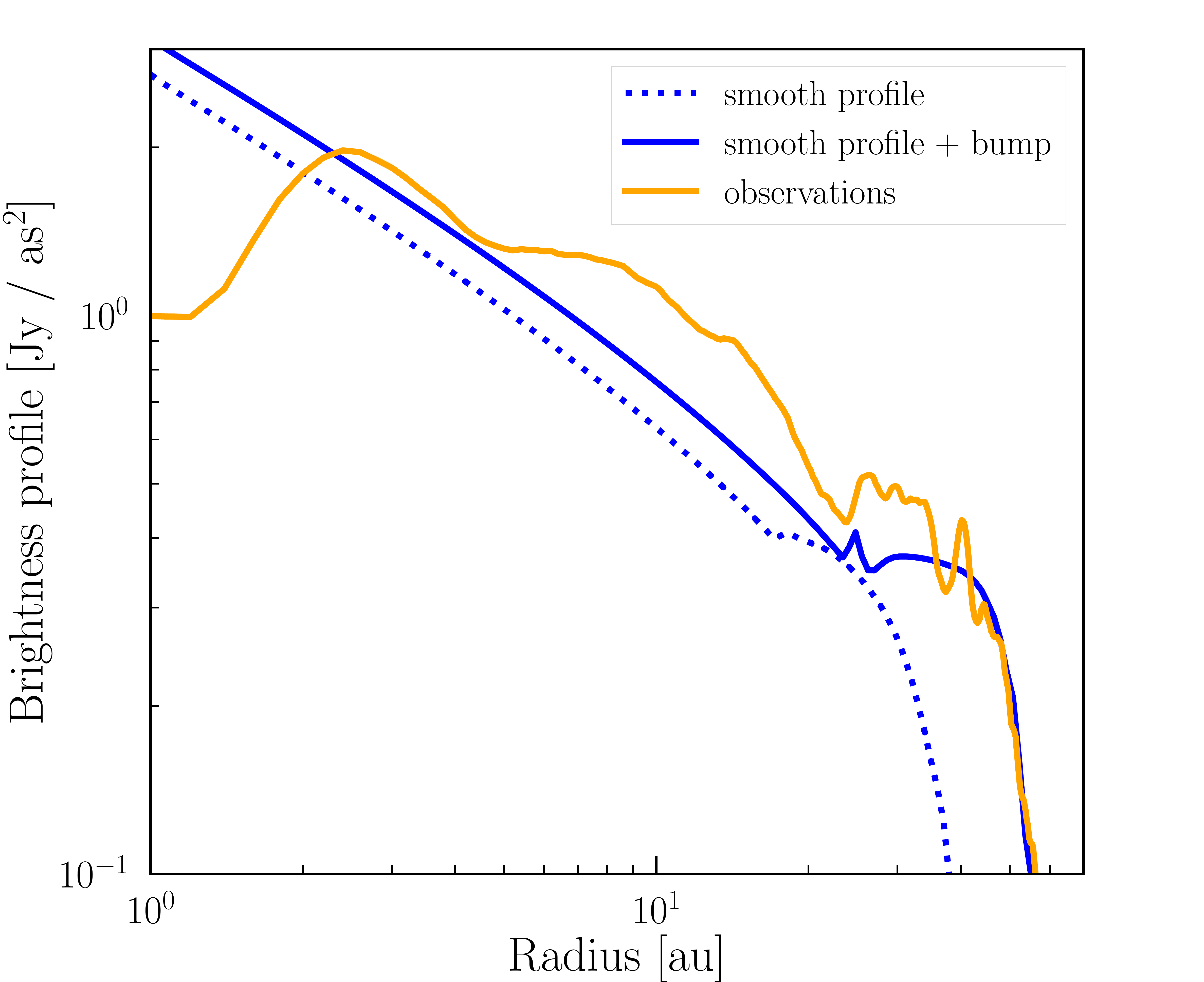}
    \caption{Comparison of the dust continuum emission at 0.87 mm as predicted by our model (blue solid line) and the observational data (orange solid line) from \cite{Andrews16}. The blue dotted line shows the model result without accounting for radial structures, highlighting their importance for this mass estimate.}
    \label{Fig:TWHya}
\end{figure}

Using this setup we evolved the TW~Hya disk and tested the model results against observational data for the dust continuum emission at 0.87 mm, from \cite{Andrews16}, as shown in Fig.\ref{Fig:TWHya}. As a proof of concept in the figure we also show the model result without including the bump located at 70 au, placing the location of the dust line $\sim 20$ au closer to the star. To fit this smooth model with the observations, we would need to assume a disk mass about $\sim 50\%$ smaller than the mass derived with the presence of the bump.

Using this model, we were able to reproduce the dust line location from observations using a disk mass of 0.11 M$_\sun$, in accord to \cite{Powell17} who estimated the same disk mass with a different combination of grain size and drift velocity, not accounting for the bump in the gas density distribution. This explains why the mass estimates are in agreement, in contrast of what one would expect from our population synthesis study, showing that the Powell expression overestimates the disk mass by a factor of $\sim 12$. This estimate is consistent with the lower limit of 0.05 M$_\sun$ given by HD measurements \citep{Bergin13, Kama16}, but not with the more recent estimate $7.7 \cdot 10^{-3} \; \mathrm{M_\sun} \leq M_{disk} \leq 2.3 \cdot 10^{-2} \; \mathrm{M_\sun}$ from \cite{Trapman17}. This new mass is up to a factor of $\sim 5$ larger than the mass derived from CO observations, depending on the model \citep{Thi10, Gorti11}, showing therefore evidence of moderate depletion of CO \citep{Powell17}.\par

\subsection{AS~209}
In this section, we test our model for a disk with strong radial substructures, AS~209. For this disk, surface density profiles are available in the literature, either from CO observations \citep{Huang16} or multiwavelength continuum observations \citep{Tazzari16}. Unfortunately high spatial resolution data of CO gas emission are not available. To model the gas structure we used the same parameterization of the dust distribution given in \cite{Fedele18}, an extensive study of the disk radial structures based on high-resolution Atacama Large Millimeter/submillimeter Array (ALMA) data. The reader should keep in mind that this is an approximation: the purpose of this section is to show a practical application of our model to a disk with prominent radial features and not an accurate modeling of the disk structure around AS~209. The gas surface density distribution is:

\begin{equation}
    \Sigma(r) = \Sigma_c \left(  \frac{r}{r_c} \right)^{-\gamma_1} \exp \left[ -\left(  \frac{r}{r_c} \right)^{2-\gamma_2} \right] \times \delta(r),
\end{equation}

where $\sigma(r)$ is the density scaling factor:

\begin{align}
    \delta(r) = 1 - A_{G1} \, \phi(R_{G1}, \sigma_{G1}) + A_{R1} \, \phi(R_{R1}, \sigma_{R1}) -\\
    A_{G2} \, \phi(R_{G2}, \sigma_{G2}) +
    A_{R2} \, \phi(R_{R2}, \sigma_{R2})
\end{align}

where the right end terms are Gaussian functions  $A \,\phi(R, \sigma)$, centered on $R$  with width $\sigma$ and amplitude $A$, representing the rings and gaps in the gas structure. The values used in this model are shown in Tab.\ref{table:AS209_params}.

\begin{table}[H]
\caption{Parameter values assumed for AS 209.}             
\label{table:AS209_params}      
\centering                          
\begin{tabular}{c c | c c}        
\hline\hline                 
Parameter & Value & Parameter & Value \\    
\hline                        
    $M_*$ & 0.9 M$_\sun$ & $R_{R1} $ & 78.65 au\\
    $M_{disk}$ & 0.11 M$_\sun$ & $\sigma_{R1}$ & 8.95 au\\
    $r_c$ & 80 au & $A_{G2}$ & 0.025\\
    $\gamma_1$ & -0.24 & $R_{G2}$ & 103.2 au\\
    $\gamma_2$ & 2.19 & $\sigma_{G2}$ & 15.6 au\\
    $A_{G1}$ & 0.03 & $A_{R2}$ & 4.8\\
    $R_{G1}$ & 61.7 au & $R_{R2}$ & 129.3 au\\
    $\sigma_{G1}$ & 8 au & $\sigma_{R2}$ & 10.5 au\\
    $A_{R1} $ & 0.80 & \\

\hline                                   
\end{tabular}
\end{table}

\begin{figure}[H]
    \centering
    \includegraphics[width=8cm]{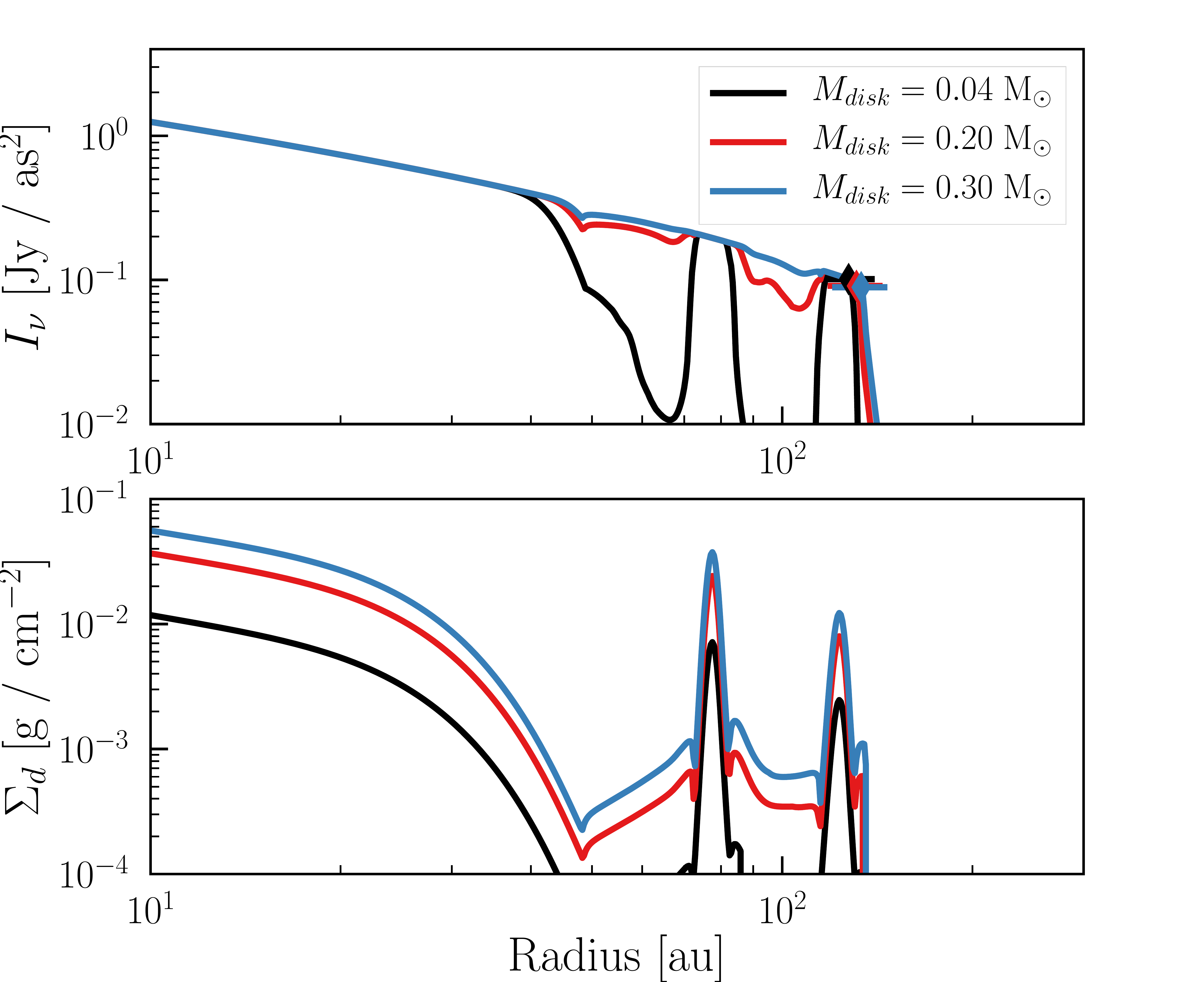}
    \caption{Comparison between the brightness profile at 0.87 mm and the density distribution of the most emitting grains.}
    \label{Fig:AS209_0.87mm}
\end{figure}

\begin{figure}[H]
    \centering
    \includegraphics[width=8cm]{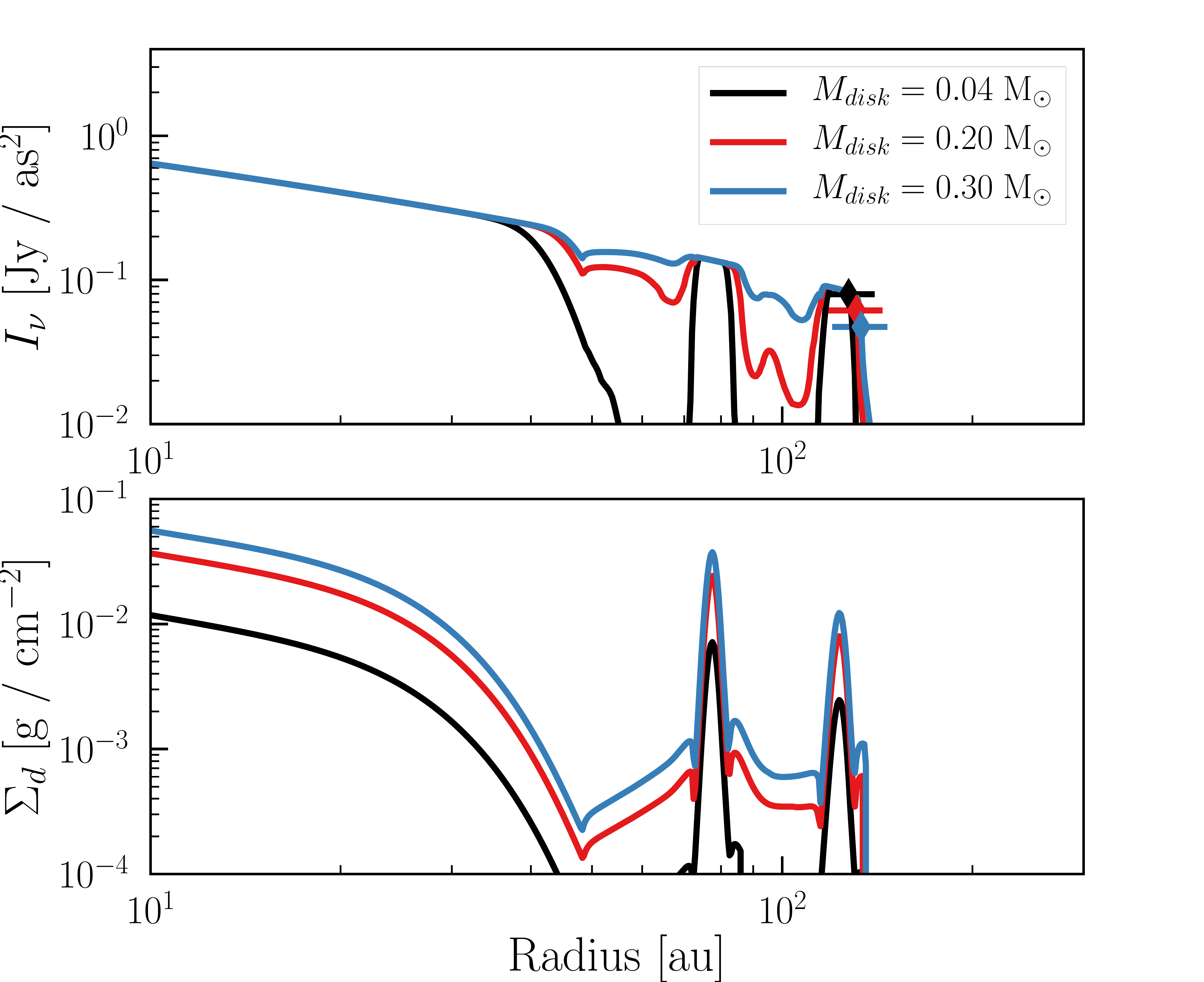}
    \caption{Comparison between the brightness profile at 1.3 mm and the density distribution of the most emitting grains.}
    \label{Fig:AS209_1.3mm}
\end{figure}

Using this parametrization we simulated the evolution of AS~209, and compared the emission at 0.87 mm wavelength with the distribution of grains dominating the emission at this wavelength in Fig.\ref{Fig:AS209_0.87mm}. The figure shows how in the case of a disk with prominent radial features, such as AS~209, grains gather at the ring locations, therefore the dust line location is tracing the trap location and lose the dependency with the disk mass. The trapping happens for grains of all sizes, therefore this result does not change with the observed wavelength as observed recently by \cite{Long20} when comparing ALMA observations at two wavelengths of protoplanetary disks with structures. In Fig.\ref{Fig:AS209_1.3mm} we see how the disk shows the same profile at 1.3 mm, and the dust line location lays at the outer bump location at both wavelengths. Multiwavelength observations can reveal when dust lines trace a bump location, that is, when the dust line cannot be used as a mass estimator. Particle trapping can affect the applicability of the mass estimate method.

\section{Summary and conclusions}
We used dust evolution models to test the applicability of the disk outer edge (dust lines) located at multiple wavelengths as a disk gas mass tracer, as proposed in \cite{Powell17, Powell19}. The dust line location is related to the maximum radial location at which grains of size $2 \pi / \lambda_{obs}$ can be observed. The robustness of this method comes from its independence on an assumed tracer abundance to derive the total gas mass. The assumption made to derive a disk mass is that the dust evolution is drift dominated, meaning that the age, growth and drift timescale are the same at the late stages of disk evolution for the grains at their dust line location. This is a reasonable assumption for evolved disks ($t_{disk} \gtrsim 1 Myr$), confirmed by previous results in the literature. At the same time, the relation between dust line location and disk mass can be dubious in disks showing radial features. These substructures can affect the grains drift timescale, and the relation between dust line location and disk mass is no longer unique.\par

To test the reliability of the dust line location as a mass estimator, we used numerical models of grain evolution to study the dust evolution in disks. We first analyzed the dependence of the dust line location on the disk mass, age, dust-to-gas ratio and stellar mass. We find that the dust line location is mainly dependent on the disk mass and the age of the system. The disk mass and age are degenerate, and to apply this method we need a reliable age estimate. Any uncertainty in the age estimate is propagated to the mass estimate.\par

Once we know the dust line to be mainly dependent on the disk mass, as the next step we calibrated its relation to the total disk mass. We applied the dust line-disk mass relation from \cite{Powell17, Powell19} to a population of synthetic disks, and while the disk masses from the models and the ones derived from the dust line are tightly correlated as expected, we found that the masses derived from the \cite{Powell17} expression are overestimated by a factor of $12 \pm 0.5$, bringing this mass estimate more in line with gas-based estimates. This result improves on the argument presented in \cite{Powell17}, based on local considerations of the grain evolution on the dust line location, by taking into account that the dust lines of different populations are not independent, as the total mass flux has to be conserved. This is not taking into account the effect of radial structures on dust evolution, and care needs to be used in comparing this result to other mass estimates. This study also demonstrates that the reliability of this mass estimate is not affected by the disk physical properties, except for a high turbulent parameter ($\gtrsim 10^{-2}$) or a very low disk mass ($\lesssim 10^{-4}$ M$_\sun$), as this would break the assumption of drift dominated dust evolution.\par

We then tested the effect of perturbations on the surface density distribution of disks: these substructures affect the dynamical timescale of grains and the location of their dust line. We find that, when these structures do not act as efficient dust traps, the dust line location moves further out, while keeping a strong dependence on the disk mass. Therefore, while the dust line-$M_{disk}$ relation depends on the disk structure, it is still a good mass estimator. When instead radial structures act as efficient particle traps, the dust line location is no longer dependent on the disk mass, and it only traces the trap location. From an observational point of view, we know when the dust line is tracing the trap location by looking at multiwavelength dust continuum observations. In case of trapping, we expect grains of all sizes to be trapped, and dust lines at different wavelength will be observed at the same location. In this case, we will need to resort to other mass estimators.\par

Lastly, to show the applicability of the method to real data, we used this technique to estimate the mass of two well studied disks: TW~Hya, characterized by an overall smooth gas structure, and AS~209, a disk featuring several gaps and traps. Our method successfully derived a mass estimate for TW~Hya of $0.11 M_\sun$, a factor of $\sim 5$ higher than CO based mass estimates. The disk in AS~209, on the other hand, is an example of when the dust line location is not an appropriate mass estimator. This disk presents evidences of particle trapping, and our model predicts a gathering of the dust lines for different dust populations at the location of the outer ring. In this case the dust line location is tracing the location of particle traps, and as such is depending not on the disk mass but on the location of the structures in the gas profile.\par

To summarize, in this paper we provide evidence showing that the dust line location is a reliable mass estimator when the dust evolution is drift dominated, although the exact relation between dust line and disk mass depends on the structure of the studied disk. This confirms the analysis of \cite{Powell17, Powell19}. This relation in turn can be calibrated by running a simple two population dust evolution model that, when there is no particle trapping in the disk, provides us the location of the dust line location for each observed dust population for a given disk mass.


%
   \begin{figure}
   \centering
         \label{FigVibStab}
   \end{figure}

\begin{acknowledgements}
The research for this paper was supported by the European Research Council under the Horizon 2020 Framework Program via the ERC Advanced Grant Origins 83 24 28.\\
DS acknowledges support by the Deutsche Forschungsgemeinschaft through SPP 1833: "Building a Habitable Earth'" (grant SE 1962/6-1).\\
T.B. acknowledges funding from the European Research Council (ERC) under the European Union's Horizon 2020 research and innovation program under grant agreement No 714769 and funding from the Deutsche Forschungsgemeinschaft under Ref. no. FOR 2634/1 and under Germany's Excellence Strategy (EXC-2094–390783311).\\
P.P. acknowledges support provided by the Alexander von Humboldt Foundation in the framework of the Sofja Kovalevskaja Award endowed by the Federal Ministry of Education and Research.\\
\end{acknowledgements}

%
%

\bibliographystyle{aa} 
\bibliography{aanda} 
\nocite{Draine03}
\nocite{Zubko96}
\nocite{Warren08}

\end{document}